\documentclass[preprint,review,12pt]{elsarticle}

\usepackage[top=2.5cm, bottom=2.5cm, left=2.7cm, right=2.7cm]{geometry}

\usepackage{fullpage}
\usepackage[english]{babel}
\usepackage[utf8]{inputenc}
\usepackage[T1]{fontenc}
\usepackage{amsmath}
\usepackage{amssymb}
\usepackage{amsfonts}
\usepackage[normalem]{ulem}
\usepackage{verbatim}
\usepackage{graphicx}
\usepackage{hyperref}
\usepackage{psfrag}
\usepackage{array} 
\usepackage{color}
\usepackage{ulem}
\usepackage{fix-cm}

\usepackage{xcolor}
\usepackage[ruled,vlined]{algorithm2e}

\newcommand{\R}{\mathbb{R}}

\newcommand{\g}[1]{\boldsymbol{#1}}
\newcommand{\PP}[1]{\left( #1 \right)}

\newcommand{\Ac}[1]{\left\{ #1 \right\}}
\newcommand{\Cr}[1]{\left[ #1 \right]}

\newcommand{\Eq}[1]{\begin{equation} #1 \end{equation}}
\newcommand{\Eqa}[1]{\begin{equation*} #1 \end{equation*}}

\newcommand{\BI}[1]{\begin{itemize} #1 \end{itemize}}

\newcommand{\COM}[1]{ }

\newcommand{\m}{^{\text{mes}}}

\usepackage{amssymb}

\graphicspath{{images/}}
\usepackage{amssymb}
\usepackage{setspace}
\usepackage{amsmath}
\usepackage{colortbl}
\usepackage{subcaption}
\usepackage{graphicx}
\newcolumntype{G}{>{\columncolor{gray!30}}c}

\begin{document}

\begin{frontmatter}



\title{Variance-based variable selection in sensor calibration with strong interferents - application to air pollution monitoring with a carbon nanotube sensor array} 


\author[lab_aff]{Marine Dumon} 
\ead{marine.dumon@univ-eiffel.fr} 
\author[lab_aff]{Bérengère Lebental}
\ead{berengere.lebental@univ-eiffel.fr} 
\author[lab_aff]{Guillaume Perrin}
\ead{guillaume.perrin@univ-eiffel.fr} 

\affiliation[lab_aff]{organization={Université Gustave Eiffel, COSYS},
            postcode={77420}, 
            city={Champs-sur-Marne},
            country={France}}

\begin{abstract}

Air and water pollution are major threats to public health, highlighting the need for reliable environmental monitoring. Low-cost multisensor systems are promising but suffer from limited selectivity, because their responses are influenced by non-target variables (interferents) such as temperature and humidity. This complicates pollutant detection, especially in data-driven models with noisy, correlated inputs. We propose a method for selecting the most relevant interferents for sensor calibration, balancing performance and cost. Including too many variables can lead to overfitting, while omitting key variables reduces accuracy. Our approach evaluates numerous models using a bias-variance trade-off and variance analysis. The method is first validated on simulated data to assess strengths and limitations, then applied to a carbon nanotube-based sensor array deployed outdoors to characterize its sensitivity to air pollutants.

\end{abstract}




\end{frontmatter}

\section{Introduction}
\label{sec:introduction}
Air and water pollution are major global public health issues \cite{PublicHealth1, PublicHealth2}. Accurate detection of hazardous pollutants requires compact, reliable, low-cost, and high-precision sensors. Sensors using innovative materials (e.g., nanomaterials, polymers, ceramics) are promising due to their high sensitivity to chemical environments \cite{milone2021gas}. However, this sensitivity also makes them prone to interference from environmental parameters. The large number of potential interferents complicates calibration, as experimental designs grow exponentially with each additional variable. This motivates in-situ calibration using real-world data for sensor adjustment \cite{delaine2019situ,ngoune2024comparison}, which exposes sensors to multiple relevant factors simultaneously. Yet, ambient variables are often highly correlated (e.g., pollutant levels vary with temperature and humidity), leading to distorted calibration laws. This limits transferability across environments and hinders interpretation. Identifying causal variables is key: omitting them leads to unexplained variance, while including too many increases cost, noise, and the risk of overfitting. 

Conventional methods for variable selection face major challenges when applied to sensor calibration in open environments. Sensitivity analysis techniques, such as Sobol indices \cite{Sobol}, Shapley effects \cite{Shapley_Cohen}, or HSIC-based methods \cite{HSIC_Veiga}, are sensitive to noise, which distorts effect estimates and leads to unreliable conclusions. Strong correlations between inputs further complicate interpretation, as non-influential variables may appear important due to their association with influential variables \cite{Iooss2019}.
Regression-based methods using statistical tests \cite{Statistical_Learning}, penalized techniques like Lasso \cite{LASSO, FUJIWARA2015367} or Ridge \cite{RIDGE}, or stepwise procedures \cite{Stepwise} assume that coefficients reflect true influence. However, this fails if unmeasured causal variables are correlated with observed ones, introducing bias.
Cross-validation, often used to assess variable relevance through predictive performance, may favor error compensation in correlated settings, masking true influences. And causal approaches (e.g., PC \cite{PC}, FCI \cite{FCI}), while allowing for prior knowledge integration, remain in practice unstable in noisy, correlated, or partially observed data, that is to say the typical conditions of open-environment calibration.

Building on prior work and its limitations, this work introduces a method for selecting only those environmental variables that are truly necessary for accurate sensor prediction, excluding non-causal or weakly influential variables. Designed to be robust and computationally efficient, the method addresses realistic challenges: highly correlated inputs, unmeasured causal factors, limited calibration data (small data), and significant noise.
The approach involves (1) generating a large set of calibration models using varying combinations of input variables, and (2) applying a probabilistic post-processing step to select models that minimize prediction variance under a parsimony constraint. Beyond identifying an optimal model, the method adopts a white-box perspective, offering indicators to assess variable sensitivities and enhance interpretability.

The paper is organized as follows: Section \ref{sec:framework} defines the problem and framework; Section \ref{sec:formalism} details the method; and Section \ref{sec:application} evaluates its performance on both simulated and experimental data.

\section{General framework}
\label{sec:framework}

We are interested in the deployment of $d$ innovative sensors in an open environment, each of which outputs a vector with $q$ components. The proposed method treats each component of the vector independently, so it is equivalent whether $d \times q $ sensors with an output of dimension $1$ are considered or $d$ sensors with an output of dimension $q$. Thus, to simplify notations, we consider a deployment of ${d_y}$ sensors with a single real output, and we gather the combined output vector in $\g{y}=(y_1,\ldots,y_{d_y})$. We then note $\g{x}\in\R^{d_x}$ as the vector gathering the concentrations of the ${d_x}$ target pollutants for which the sensor was designed. Whether by design or not, the sensors may react to other environmental variables (e.g., temperature, relative humidity, other pollutants) in addition to the target chemicals. These interfering variables are divided into two categories: we denote by $\g{z}\in\R^{d_z}$ the measured environmental variables (e.g., those measured by instruments separate from the sensors being calibrated), and by $\g{u} \in \R^{d_u}$ the environmental variables that are not measured (either because it is not technically possible or simply because the sensitivity of the sensors to them was not been anticipated). Since the goal of this work is to select influential variables, the formalism allows for the possibility that the sensor outputs are in fact sensitive to only a subset of the components of $\g{z}$. This leads to the general model formulation:
\Eq{\label{Eq.general}\g{y} = \mathcal{M}(\g{x},\g{z}, \g{u}), }

\noindent{}where $\mathcal{M}$ is an unknown function, hereafter called the \emph{calibration} function. We assume that the quantities $\g{x}$, $\g{z}$ and $\g{y}$ (but not $\g{u}$) were measured at $n$ different times $i$. These measured values, which are assumed to be affected by noise and therefore slightly different from the true values, are noted $\g{x}_i^\text{mes}$, $\g{z}_i^\text{mes}$ and $\g{y}_i^\text{mes}$ respectively, and gathered in $\mathcal{D}_n = \{ \g{x}^\text{mes}_i, \g{z}^\text{mes}_i, \g{y}^\text{mes}_i \} _{i=1}^n$. 

Sensor deployment involves two stages: calibration, where an approximation of $\mathcal{M}$ is built from $\mathcal{D}_n$, and prediction, where pollutant concentrations are estimated from sensor outputs and environmental variables.
Calibration in open environments presents several challenges. First, $\mathcal{M}$ is typically non-linear with complex dependencies between $\g{x}$ and $\g{z}$, requiring flexible models. Second, measurement noise increases the risk of overfitting, requiring robust statistical methods. Third, irrelevant but correlated variables may bias the model. Fourth, unmeasured variables may also affect the sensor response. Finally, calibration typically occurs under a small data regime, limiting the use of high-capacity models such as neural networks.
Addressing these challenges - open-environment data, non-linearity, noise, unobserved inputs, small datasets, and variable interdependence - is central to this work. Subsections \ref{subsec:Model_construction} through \ref{subsec:Post_processing} describe our approach to select the most influential components of $\g{z}$, denoted $\g{z}_{\g{\alpha}}$.
In the prediction phase, the pollutant concentrations at time $t$ are estimated using the obtained approximation of $\mathcal{M}$ and the noisy observations of $\g{z}_{\g{\alpha}}$ and $\g{y}$. As discussed in \cite{SMAI_Calibration}, this step remains difficult under the constraints listed above (noise, small data, imperfect models). Our adapted solution is described in Section \ref{subsec:pollutant_concentration}.

\paragraph{Notations} 
For any time $\star$ not included in the calibration data set $\mathcal{D}_n$ (which, as a reminder, is indexed by $i$), we define $\g{z}^\text{mes}_\star$ and $\g{y}^\text{mes}_\star$ as the measured values of $\g{z}$ and $\g{y}$. Since $\g{x}_\star$ is not observed, the role of the sensor is to predict it. We also define the random vectors $(\g{X}\m, \g{Z}\m_{\g{\alpha}})$ to represent possible values of $\g{x}$ and $\g{z}_{\g{\alpha}}$ at any time $\star$. These vectors incorporate two sources of variability: measurement uncertainty at a given time, and natural fluctuations over time. They play a key role in the variable selection process described in Section \ref{subsec_varSelect}.


\section{Proposed formalism for the calibration and prediction phases}
\label{sec:formalism}

\subsection{Model definition}
\label{subsec:Model_construction}
An approximate calibration model is built independently for each component of $\g{y}$. For ease of reading, we therefore introduce in Sections \ref{subsec:Model_construction} to \ref{subsec:Post_processing} the variable $v\in\R$ to denote a specific component of $\g{y}$: $v_i$ is the true value of $v$ at time $i$, and $v\m_i$ is its measured value. 
We propose to use the formalism of the generalized linear regression (GLR) to construct an approximation of the calibration model. This choice is motivated by the relative robustness of GLR to noise, its ease of interpretation, and the fact that it can be applied in the small data context. More specifically, we propose to approximate the link between $v\m_i$, $\g{x}\m_i$, and $\g{z}\m_i$ in the following form

\Eq{
{v}\m_{i} = \g{f}(\g{x}\m_i, \g{z}\m_i)^T \g{\beta} + {\varepsilon}^{\text{mod}}_{i},
 \label{modelRegreLin}
}

\noindent{}where $\g{\beta}$ is a $k$-dimensional vector to be estimated, 
$\g{f}$ is a selected vector that includes $k$ different polynomial functions (with cross-terms) of $\g{x}\m_i$ and $\g{z}\m_i$ of total order less than or equal to $p$,
and 
$\varepsilon^{\text{mod}}_{i}$ is a model error term (to be estimated) to account for the error in approximating the calibration model, the measurement error in $v_i$, and the potential contributions of the unmeasured variables $\g{u}_i$. 
The choice of these $k$ functions is one of the contribution of this work, and will be at the heart of the discussions in Section \ref{subsec_varSelect}.

Given $\g{v}\m=({v}\m_1,\ldots,{v}\m_n)$ the vector formed by all the measurements of ${v}$, and using a least-squares approach \cite{Least_square}, the vector $\g{\beta}$ can be estimated by:

\Eq{\widehat{\g{\beta}} = \PP{{\g{H}^\text{mes}}^T\g{H}^\text{mes}}^{-1}{\g{H}^\text{mes}}^T\g{v}\m, \label{def_theta_hat}}

\noindent{}where $\g{H}^\text{mes}$ is a $(n\times k)$-dimensional matrix such that for each $1\leq i\leq n$ and each $1\leq j\leq k$,
\Eq{\PP{\g{H}^\text{mes}}_{ij} = f_j(\g{x}\m_i, \g{z}\m_i). }

It is important to note that Eq. (\ref{modelRegreLin}) links the values of $\g{x},\g{z},\g{y}$ that are in $\mathcal{D}_n$, but that such an expression only makes sense if it is also valid for linking $\g{x}_\star,\g{z}_\star,\g{y}_\star$, i.e., for linking the values of $\g{x},\g{z},\g{y}$ at times not contained in $\mathcal{D}_n$. In this case, the prediction of the measured value of $v$ at any time $\star$ can be written

\Eq{
\widehat{{v}}(\g{x}\m_\star,\g{z}\m_{\star}) = \g{f}(\g{x}\m_\star,\g{z}\m_{\star})^T \g{\widehat{\beta}} + \widehat{\theta}\xi, \ \ \widehat{\theta}^2 =\frac{1}{n-k} \sum_{i=1}^n \PP{{v}\m_{i} - \g{f}(\g{x}\m_i, \g{z}\m_i)^T\widehat{\g{\beta}}}^2,
\label{modelRegreLin_star}
}

\noindent{}where $\xi$ is a normalized model error term. As is common, a probabilistic approach is adopted, and the model error term $\xi$ and the measurement errors are considered to be random. Since the estimators $\widehat{\g{\beta}}$ and $\widehat{\theta}$ are functions of the measurements, they are also random, and we can assign them a mean and a variance. Let $\g{m}_{\beta}$ and $\g{C}_{\beta}$ be the mean and covariance matrix of $\widehat{\g{\beta}}$, and $\theta^2$ be the expectation of $\widehat{\theta}^2$. Note that the quantities $\g{m}_{\beta}$, $\g{C}_{\beta}$ and $\theta^2$ are \textit{a priori} unknown.
For reasons of identifiability, we then assume (1) that the measurement errors and the model errors are statistically independent of each other, so that $\xi$ is independent of $\widehat{\g{\beta}}$ and $\widehat{\theta}$, and (2) that the model error has variance $1$ (the amplitude of the model error is then completely carried by $\widehat{\theta}$) and mean $0$ (this last assumption is almost always true if a constant term is properly included in the components of $\g{f}$). 

\subsection{Variable selection}
\label{subsec_varSelect}
This section presents the core of the proposed method, which aims to select the components of $\g{z}$ that most improve the accuracy of pollutant concentration estimates. The approach relies on the analysis of how the model prediction variance evolves when variables are included or excluded, allowing for a more precise identification of the truly influential inputs.
Rather than applying brute-force variance minimization, the method accounts for the fact that, in highly correlated settings, adding variables—even non-influential ones—may slightly reduce variance due to shared information. However, such gains are often marginal compared to the increase in model complexity. Thus, the goal is to strike a balance between minimizing variance and maintaining parsimony. This trade-off is handled by using the Bayesian Information Criterion (BIC) \cite{Priestley1981}, which penalizes unnecessary complexity.

Let us formalize the process here: for any non-empty vector of indices $\g{\alpha}\subset \Ac{1,\ldots,d_z}$, $\g{z}_{\g{\alpha}}$ is the vector made up of the components $z_\ell$ of $\g{z}$ such that $\ell\in\g{\alpha}$. Now suppose we try to predict the value of $\g{x}$ at any time that is not part of the calibration set $\mathcal{D}_n$. 
To do this, we will only have measurements of $\g{y}$ and $\g{z}$.
The most likely value of $\g{x}$ from the sensor's point of view will therefore be the value such that the prediction of $\g{y}$ knowing $\g{x}$ and $\g{z}$ will be close to its available measurement. And the more accurate this predictor is, in the sense that the smaller its prediction variance, the more accurate we can expect the estimate of $\g{x}$ to be. To formalize this, we reuse the notations $\g{X}\m$ and $\g{Z}\m_{\g{\alpha}} $ introduced at the end of Section \ref{sec:framework}. Recall that $\g{X}\m$ and $\g{Z}\m_{\g{\alpha}} $ are two random vectors whose fluctuations are supposed to characterize the measured values of $\g{x}$ and $\g{z}_{\g{\alpha}}$ that the sensor is likely to encounter during its deployment. Following Eq. (\ref{modelRegreLin_star}), the predictor of each component $v$ of $\g{y}$ at $(\g{X}\m,\g{Z}\m_{\g{\alpha}})$ can be written as

\Eq{\widehat{{v}}(\g{X}\m,\g{Z}\m_{\g{\alpha}}) = \g{f}(\g{X}\m,\g{Z}\m_{\g{\alpha}})^T \g{\widehat{\beta}} + \widehat{\theta}\xi.
\label{def_y_pred}}

Note that $\g{f}$, $\g{\widehat{\beta}}$, and $\widehat{\theta}$ implicitly depend on the choice of $\g{\alpha}$, and could have been written with a subscript $\g{\alpha}$, which we chose not to do so as not to overload the mathematical expressions to come. From a mathematical point of view, finding the values of $\g{\alpha}$ and $\g{f}$ that lead to a minimum \emph{variance}, once averaged over all the possible values of $\g{x}$ and $\g{z}_{\g{\alpha}}$, amounts to solving the following optimization problem:

\Eq{\g{\alpha}^{\text{opt}}\in \arg\min_{\g{\alpha}} \mathbb{E}_{\g{X}\m,\g{Z}\m_{\g{\alpha}}}\Cr{\text{Var}\PP{\widehat{{v}}(\g{X}\m,\g{Z}\m_{\g{\alpha}}) \vert \g{X}\m,\g{Z}\m_{\g{\alpha}}}}, \label{defPbOptim}}

\noindent{}where the notation $\mathbb{E}_{\g{X}\m,\g{Z}\m_{\g{\alpha}}}\Cr{\cdot}$ is introduced to clearly indicate that the average is taken over the random variables $\g{X}\m$ and $\g{Z}\m_{\g{\alpha}}$. This expectation of conditional variance is noted $V(\g{\alpha},\g{f})$ in the following, and it rewrites as (see \ref{appendixA} for details)
\Eq{V(\g{\alpha},\g{f}) = \mathbb{E}_{ \g{X}\m,\g{Z}\m_{\g{\alpha}} } \Cr{ \text{Var}\PP{ \widehat{{v}}(\g{X}\m,\g{Z}\m_{\g{\alpha}}) \vert \g{X}\m,\g{Z}\m_{\g{\alpha}} }}
 = \g{m}_{f}^T\g{C}_{\beta}\g{m}_{f} + \theta^2 + \text{Tr}\PP{\g{C}_f\g{C}_\beta},
}

\noindent{}where $\g{m}_{f}=\mathbb{E}_{\g{X}\m,\g{Z}\m_{\g{\alpha}}}\Cr{\g{f}(\g{X}\m,\g{Z}\m_{\g{\alpha}})}$, $\g{C}_f=\text{Cov}_{\g{X}\m,\g{Z}\m_{\g{\alpha}}}\PP{\g{f}(\g{X}\m,\g{Z}\m_{\g{\alpha}})}$, $\theta^2=\mathbb{E}[\widehat{\theta}^2]$, and $\g{C}_{\beta}=\text{Cov}(\widehat{\g{\beta}})$. The three terms in this expression can be interpreted as follows: (1) $\g{m}_{f}^T\g{C}_{\beta}\g{m}_{f}$ is linked to the uncertainty in estimating $\g{\beta}$ due to the finite and noisy nature of the training data set, and is expected to tend to $0$ as the size of the data set $n$ tends to infinity; (2) $\theta^2$ characterizes the model error and is expected to decrease with increasing model complexity; (3) $\text{Tr}\PP{\g{C}_f\g{C}_\beta}$ is associated with the potential lack of robustness of an overly complex model, and unlike $\theta^2$, is expected to increase with increasing model complexity. 

The terms $\g{m}_{f}$, $\g{C}_{\beta}$, $\mathbb{E}[\widehat{\theta}^2]$, and $\g{C}_f$ are unknown, since they depend through $\g{X}\m$ and $\g{Z}\m_{\g{\alpha}}$ on unknown fluctuations of $\g{x}$ and $\g{z}$. Thus, we propose to approximate them based solely on the calibration data set $\mathcal{D}_n$. First, we estimate $\g{m}_{f}$ and $\g{C}_f$ using a classical Monte Carlo approach, so that:
\Eq{\hspace{-1cm}\g{m}_{f}\approx \widehat{\g{m}}_{f}=\frac{1}{n}\sum_{i=1}^n \g{f}_i, \ \ \g{C}_f\approx \frac{1}{n-1}\sum_{i=1}^n \PP{\g{f}_i-\widehat{\g{m}}_{f}}\PP{\g{f}_i-\widehat{\g{m}}_{f}}^T, \ \ \g{f}_i=\g{f}(\g{x}\m_i, \g{z}\m_{i,\g{\alpha}}).}

Estimating $\g{C}_{\beta}$ and $\mathbb{E}[\widehat{\theta}^2]$ is less straightforward, since the estimators $\widehat{\g{\beta}} $ and $\widehat{\theta}^2$ are functions of the complete set of points in $\mathcal{D}_n$, and several sets of $n$ points with the same properties as $\mathcal{D}_n$ would  be needed to compute their statistics. To get around this difficulty, we use a bootstrap approach
\cite{Bootstrap}: we select $n$ points in $\mathcal{D}_n$ independently and with replacement (e.g., a single point can be retained several times), then we calculate the associated values of $\widehat{\g{\beta}}$ and $\widehat{\theta}$. We repeat this procedure a large number of times, from which we derive empirical approximations of $\g{C}_{\beta}$ and $\mathbb{E}[\widehat{\theta}^2]$, again using the classical Monte Carlo approach. 

In this search for a compromise between complexity and variance, it is also important to note that the choice of $\g{f}$ is central, since the total polynomial order of the model and the number of cross terms contribute much more to the complexity of the model than the number of variables. Therefore, the optimal value for $\g{\alpha}$ should actually be sought as the solution of
\Eq{\min_{\g{\alpha}} \min_{\g{f}\in \mathcal{P}(p,\g{\alpha})} V(\g{\alpha},\g{f}),}

\noindent{}with $\mathcal{P}(p,\g{\alpha})$ the set of vector-valued functions whose components are polynomial functions with total degree less than $p$. This minimization with respect to both $\g{\alpha}$ and $\g{f}$ introduces a new difficulty. In fact, the minimization with respect to $\g{\alpha}$ only can be done in a purely combinatorial manner, e.g., brute-force testing all the $2^{d_z}$ combinations of $\g{\alpha}$ to select the best $\g{\alpha}$, since the number of measured environmental variables is usually limited (typically $d_z \leq 10$). Conversely, even reasonable polynomial orders (in the applications below, $p$ is chosen equal to 3) lead to an unreasonable number of  models to be tested in a combinatorial approach (this number would be equal to $2^{\binom{d_x+d_z+p}{p}}$, which would give a number of combinations of the order of $10^{17}$ for $d_x=2$, $d_z=3$ and $p=3$). 

As an alternative, a \emph{greedy} approach is proposed here. For each value of $\g{\alpha}$, we first choose a value for $p$. This allows us to define $\g{f}^{\text{full}}$ as the particular element of $\mathcal{P}(p,\g{\alpha})$ that contains all the polynomial functions of total order less than $p$ that are a function of $\g{x}$ and $\g{z}_{\g{\alpha}}$. By construction, $\g{f}^{\text{full}}$ is a vector gathering $\binom{d_x+\sharp \g{\alpha}+p}{p}$ polynomial functions, where $\sharp \g{\alpha}$ is the number of elements of $\g{\alpha}$. To make the values of the projection coefficients comparable, except for the term corresponding to the constant function equal to $1$, we then normalize the components of $\g{f}^{\text{full}}$ by removing their empirical means and dividing by their empirical standard deviations, both computed from $\mathcal{D}_n$. Of course, other types of normalization could have been considered. We then evaluate the prediction variance for $\g{f}^{\text{full}}$ and for a series of functions $\g{f}^{\text{partial}}$ derived from $\g{f}^{\text{full}}$ by removing each of its components one by one, starting with the term associated with the lowest $\g{\beta}$ coefficients. By repeating this procedure for each value of $\g{\alpha}$, we obtain a large number of forms for $\g{f}$, and we finally concentrate on the pair $(\g{\alpha},\g{f})$ {that minimizes } $V(\g{\alpha},\g{f})$. 

Nevertheless, such a procedure often leads to the selection of the \emph{complete model}, i.e., the model that includes all the environmental variables and all the polynomial terms (for a given polynomial order), which is neither optimal nor satisfactory. This issue is solved by penalizing complex models: instead of directly minimizing $V(\g{\alpha},\g{f})$, one minimizes its BIC version according to the formula

\Eq{\text{BIC}(\g{\alpha},\g{f})=n\times \log(V(\g{\alpha},\g{f})) + k(\g{f})\times \log(n),}

\noindent{}where $k(\g{f})$ is the number of polynomial functions that are in $\g{f}$, and $n$ is the number of points in $\mathcal{D}_n$. 

In summary, it is proposed to choose the optimal value of $\g{\alpha}$ as the one that minimizes the BIC criterion when testing a large number of polynomial models (using a greedy removal procedure) for the $2^{d_z}$ different configurations of the components of $\g{z}$. It is worth noting that other criteria besides the BIC have been tested to automatically achieve the trade-off between model variance and complexity, e.g. by comparing the variance with the measurement noise of the variables, but the BIC proved to be the most reliable in our numerical tests.

\paragraph{Remark}
During variable selection, $\g{x}$ and $\g{z}$ are treated symmetrically, so the method can be applied to both $\g{x}$ and $\g{z}$. The only constraint is the total number of variables, which must remain tractable for combinatorial exploration.
This point is practically relevant because sensor sensitivities often change significantly between laboratory and field conditions, so the actual targets and the actual interferents may not be known exactly. Consequently, an initial \emph{agnostic} interpretation phase is required, i.e., all variables are initially assumed to be interferents. The proposed variable selection protocol is then used to identify which variables truly define the sensor's targets. This agnostic approach is demonstrated in Section \ref{sec:application} using data from a novel carbon nanotube sensor array.

\subsection{Post-processing of the different tested models}
\label{subsec:Post_processing}

The method in Section \ref{subsec_varSelect} primarily returns the optimal subset $\g{z}_{\g{\alpha}}$ and its associated model $\g{f}$. In addition, for any number $\ell$ of selected variables (from 1 to $d_z$), it also returns the best performing combination of variables and the corresponding variance. Plotting $\ell$ against this variance forms a Pareto front \cite{Pareto_front}, highlighting optimal tradeoffs between prediction accuracy and model simplicity. 
For each $\ell$, the index vector $\g{\alpha}^{\text{opt}}_\ell$ identifies the best subset of components of $\g{z}$ that minimizes the BIC. The evolution of the selected variables as $\ell$ increases provides insight into their relative influence. For example, the most influential variable is expected to appear first, followed by the next most influential, and so on. This visualization, further illustrated in Section \ref{sec:application} Figure \ref{fig:Pareto_PME_specific_case}) is a powerful tool for interpreting sensor sensitivities.

This simple ranking between variables is helpful to interpret the sensor sensitivities qualitatively, but not quantitatively. Moreover, it fails to interpret variable exchanges between steps (e.g., the variables of step $\ell+1$ do not include all the variables of step $\ell$), a situation that has sometimes been observed in experimental data. To provide a more quantified and reliable interpretation of variable influence, a method of variance decomposition between variables (including the components of $\g{x}$ and $\g{z}$ and the model error) is now introduced. Classical Sobol indices \cite{Sobol} would be the natural tool here, but they rely on an assumption of statistical independence between inputs that does not hold in our context. When dependencies exist, as is the case between $\g{x}$ and $\g{z}$, the Shapley indices \cite{Shapley_Cohen}, and more recently the indices based on Proportional Marginal Effects (PME) \cite{PME} provide suitable alternatives. In this paper, we have chosen to use the PME indices {(see \ref{appendix_C} for their mathematical definition)} because of the following properties: 

\BI{
\item Positivity and Interpretability: PME values are non-negative and add up to the total model variance.
\item Causal Relevance: A variable with no causal effect will have a PME of zero, even if it is statistically correlated with the output.
\item Influence Hierarchy: A high PME value corresponds to a strong influence on the model output, in the sense that the proportion of variance attributed to it is high.
}

These features make PME indices particularly suitable for our purposes. In practice, calculating these indices requires the availability of a model linking sensor inputs and outputs, which is precisely what the method developed here aims to build. Therefore, the PME indices are computed after model construction to improve interpretability and to reliably assess the influence of variables.

\paragraph{Remark}
As we are considering noisy calibration data, the overall method can be further improved by again using a bootstrap approach. This consists in repeating the entire procedure $m$ times on $m$ different data sets, all constructed by selecting $n$ points of $\mathcal{D}_n$ independently and with replacement. It is then possible to compute statistics on the results obtained, for the PMEs indices, for the Pareto front, and for the optimal vector $\g{z}_{\g{\alpha}}$. 

\paragraph{Remark} 
An additional post-processing of the optimal model is proposed in Section \ref{Appendix_D}, aiming to compute the \emph{resolution} of the sensor to its targets $\g{x}$ and interferents $\g{z}_{\g{\alpha}}$.

\subsection{Estimation of the pollutant concentration}
\label{subsec:pollutant_concentration}
In addition to its ability to properly identify influential variables, the proposed method should be evaluated for its ability to provide a reliable estimate of $\g{x}_{\star}$ when using the optimal model for each sensor output. By specifying the subscripts $m\in\Ac{1,\ldots,d_y}$ for each quantity associated with each sensor output, the link between sensor inputs and outputs presented in Eqs. (\ref{modelRegreLin}) and (\ref{def_y_pred}) can now be written in vector form as:

\Eq{\g{y}\m_\star= \g{F}(\g{x}\m_{\star}, \g{z}\m_\star)\widehat{\g{\beta}}^{\text{tot}} + \widehat{\g{\Theta}}\g{\xi},\label{pred_vectoriel}}

\noindent{}with $\widehat{\g{\beta}}^{\text{tot}}=(\widehat{\g{\beta}}_1,\ldots,\widehat{\g{\beta}}_{d_y})$, $\g{\xi}=(\xi_1,\ldots,\xi_{d_y})$, $\g{F}=\text{diag}(\g{f}_1,\ldots,\g{f}_{d_y})$ and $\widehat{\g{\Theta}}=\text{diag}(\widehat{\theta}_1,\ldots,\widehat{\theta}_{d_y})$, where $\text{diag}(\cdot)$ is the operator that transforms a set of vectors or matrices into a block diagonal matrix where each diagonal block corresponds to the elements of the considered set of vectors or matrices. 
It is important to remember here that the optimal polynomial forms and the optimal vectors $\g{\alpha}_m^{\text{opt}}$ are derived independently for each sensor output $m$, and thus are expected to be different from each other. To be completely accurate, the formula should thus contain a reference to $\g{\alpha}_m^{\text{opt}}$. For the sake of readability, however, we simplify the notation again and refer only to $\g{z}\m_\star$ in this section, leaving implicit the fact that only certain components of this vector are actually retained. 

From a Bayesian point of view, consistent with the probabilistic formalism introduced earlier, we model $\g{x}_\star$ as a random vector and search for the value that maximizes its posterior probability density function (PDF). Following the work achieved in \cite{SMAI_Calibration}, this posterior PDF can be approximated in closed form as (see Section \ref{appendix_B} of Section \ref{Appendix_D} for the expressions of $\g{m}$ and $\g{C}$) 

\Eq{\widehat{f}_{\g{x}_\star \vert \g{z}\m_\star,\g{y}\m_\star}(\g{x})= c\times  \frac{\exp\PP{-\frac{1}{2}\PP{\g{y}\m_\star- \g{m}(\g{x})}^T\g{C}(\g{x})^{-1}\PP{\g{y}\m_\star- \g{m}(\g{x})}}}{\text{det}\PP{\g{C}(\g{x})}^{1/2}} \times f_{\g{x}_\star \vert \g{z}\m_\star}(\g{x}), \label{PDF_post_x}
}

\noindent{}where $c$ is a constant and $f_{\g{x}_\star\vert \g{z}\m_\star}$ is the PDF of $\g{x}_\star$ given $\g{z}\m_\star$, whose role is to ensure the best possible transferability of calibration results: When the sensor is used in the calibration environment, it can be approximated by $f_{\g{x}\m_\star \vert \g{z}\m_\star}$ using the points in $\mathcal{D}_n$ and relying on any statistical inference technique. In Section \ref{sec:application}, we will focus on the Kernel Density Estimation (KDE) for such constructions \cite{perrinCSDA2017,SMAI_Calibration}. 
Conversely, if the sensor is to be used in a different environment, this should be elucidated separately for that new environment (see \cite{Marin2007} for more details on expert elicitation). Finally, the value of $\g{x}$ that maximizes $\widehat{f}_{\g{x}_\star \vert \g{z}\m_\star,\g{y}\m_\star}(\g{x})$ defines an estimator of $\g{x}_\star$, and confidence intervals can be proposed for each component of $\g{x}_\star$ by extracting quantiles from this approximated posterior PDF.

\section{Application}
\label{sec:application}

The proposed method is evaluated on both simulated and experimental data sets. Simulated data, detailed in Section \ref{subsec:application_simulated}, provide a controlled setting to test the method’s ability to identify influential variables and to assess its robustness in complex scenarios. Then, in Section \ref{subsec:experimentalDataset}, the method is applied to real outdoor measurements. In this second case, the influential variables are unknown, and the data are affected by uncontrolled variations. 

In the rest, the expression \emph{simple model} denotes the model that includes all the components of $\g{x}$ as inputs and no component of $\g{z}$. The term \emph{selected model} corresponds to the model selected by the proposed method, while the \emph{complete model} includes all the components of $\g{x}$ and $\g{z}$ (even if they are not influential). 

\subsection{Application on the simulated data set}
\label{subsec:application_simulated}

The synthetic data set includes the main features generally observed in experimental data sets for multisensor systems deployed in an open periodic environment \cite{ngoune2024comparison, Ozone_CO_IEEE_proce}. In this section, only one sensor is considered ($d_y=1$), and we focus on the case $d_x=1$, $d_z=5$ and $d_u=1$. The variables $x$, $\g{z}$ and $u$ are simulated as centered Gaussian random variables with a covariance matrix $\g{C}(\rho,\rho_u)$ parameterized by two constants $\rho,\rho_u\in[0,1)$ so that:

\Eq{\setstretch{1}\g{C}(\rho,\rho_u)=\text{Cov}((x,\g{z},u),(x,\g{z},u))= \Cr{
\begin{array}{ccccccc}
    1 & \rho & -\rho & \rho & \rho & -\rho & \rho_u \\
    \rho & 1 & -\rho & \rho & \rho & -\rho & \rho_u \\
    -\rho & \rho & 1 & -\rho & -\rho & \rho & \rho_u \\
    \rho & \rho & -\rho & 1 & \rho & -\rho & \rho_u \\
    \rho & \rho & -\rho & \rho & 1 & -\rho & \rho_u \\
    -\rho & -\rho & \rho & -\rho & -\rho & 1 & \rho_u \\
    \rho_u& \rho_u & \rho_u & \rho_u & \rho_u & \rho_u & 1 
\end{array}
}.}

The relationship between these inputs and the sensor output is defined by the following relationship:\Eq{
\begin{split}
     y & = \frac{3}{2}\text{log}(x+4) +  \text{atan}\PP{\frac{z_1}{2}} \PP{1+0.1\frac{z_1}{2}} +
     \text{atan}\PP{\frac{z_2}{2}} \PP{1+0.1\frac{z_2}{2}} \\
     & + 0.1\text{atan}\PP{\frac{z_3}{2}}\PP{1+0.1\frac{z_3}{2}} 
      + 3\text{cos}\PP{\frac{z_1+z_3}{6}} + 2\text{cos}\PP{\frac{x + z_2}{6}} + \alpha_u u. 
\end{split}
\label{modelfonctionanaly}
}

\begin{figure}
    \centering
    \includegraphics[width=0.99\linewidth]{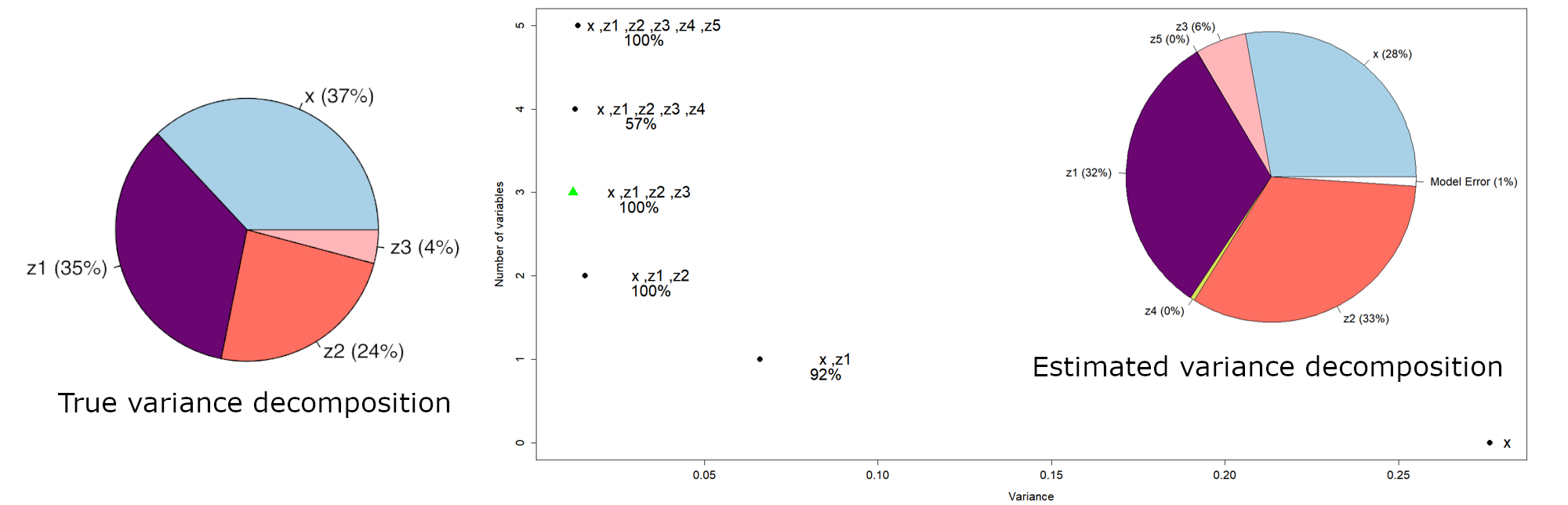}
    \caption{Left - PME-based variance decomposition of the sensor output computed using the real model, for $(\sigma\m,\rho,\alpha_u)=(5\%,0.8,0)$. Right - Average Pareto front computed over 100 bootstrap iterations, for $(\sigma\m,\rho,\alpha_u)=(5\%,0.8,0)$, illustrating the tradeoff between the number of variables selected and the associated prediction variance. Each point represents the average over the 100 replicates of the optimal variance for the corresponding number of variables. The variables listed next to each point are the most frequently selected variables across the 100 replicates, with the percentage below indicating the frequency of occurrence. The green triangle marks the optimal subset of variables without the bootsrap procedure, regardless of the number of variables. On the right, the average PME-based variance decomposition obtained for the optimal model is shown, featuring an almost perfect match with the reference.}
    \label{fig:Pareto_PME_specific_case}
\end{figure}

This relationship is non-linear and includes cross effects. It introduces a clear hierarchy between the input variables: the variables $x$ and $z_1$ are the most influential ones on $y$, then comes $z_2$ with a slightly smaller influence, and finally $z_3$ with an even smaller one (see Figure \ref{fig:Pareto_PME_specific_case}left). The constant $\alpha_u$ is used to modulate the influence of the uncontrolled variable $u$. The variables $z_4$ and $z_5$ do not appear in the model. Thus, in terms of variable selection, we want the algorithm to select only $z_1$, $z_2$ and $z_3$, regardless of the values of $\rho$, $\rho_u$ and $\alpha_u$, and regardless of the measurement noise. 

For each numerical experiment, two data sets with the same features were generated: one was used as a training data set for variable selection (calibration step), while the other served as a test data set for pollutant concentration estimation (inversion step). In this first application, we chose a training set of size $n=200$. The values of $x,\g{z},y$ were perturbed by a Gaussian noise with standard deviation $\sigma\m$. 

\subsubsection{How the method works on a specific example}
\label{subsubsec:Specific_example}

In this section, we focus on the case $\alpha_u=0$ (no unobserved variable), $\sigma\m=5\%$ (significant level of noise on the inputs and on the outputs) and $\rho=0.8$ (strong correlation between the inputs). The results below discuss first the accuracy of the variable selection, then the accuracy of the estimation of $x$. 

\begin{table}[!ht]
    \centering
    \begin{tabular}{|ccccc|}
    \hline
          $z_1$ &$z_2$ & $z_3$ & $z_4$ & $z_5$      \\
         \hline
          100 & 100 & 100 &8&10\\

         \hline
    \end{tabular}
    \caption{Percentage of selection of each variables over the 100 repetitions, for $(\sigma\m,\rho,\alpha_u)=(5\%,0.8,0)$. Variables $z_1$, $z_2$, and $z_3$ are always selected, $z_4$ and $z_5$ rarely.}
    \label{tab:simulated_robust_specific case}
\end{table}

The variable selection results are summarized in Table \ref{tab:simulated_robust_specific case} and illustrated in Figure \ref{fig:Pareto_PME_specific_case}. Table \ref{tab:simulated_robust_specific case} reports how frequently each variable $z_1$ through $z_5$ is selected over 100 bootstrap iterations using the BIC-based criterion introduced in Section \ref{subsec_varSelect}. The results show that $z_1$, $z_2$, and $z_3$ are consistently selected in 100$\%$ of the time, while $z_4$ and $z_5$ are rarely selected. This outcome is in line with expectations: although $z_4$ and $z_5$ do not have a direct effect on $y$, their strong statistical correlation with the influential variables ($z_1,z_2,z_3$) can still lead to their occasional selection. Measurement noise and the approximation error of the polynomial model may also contribute to this behavior.

Figure \ref{fig:Pareto_PME_specific_case} provides deeper insight into this selection process by showing a Pareto front of the optimal subsets of increasing size, as defined in Section \ref{subsec:Post_processing}. This allows us to rank the variables by influence: $z_1$ is selected first in 92$\%$ of cases (with $z_2$ chosen first in the remaining 8$\%$), followed by the systematic inclusion of $z_2$, then $z_3$, in the 2- and 3-variable subsets. The inclusion of each of these 3 variables leads to a significant reduction of the prediction variance. Conversely, the inclusion of $z_4$ or $z_5$  in the 4- and 5-variable subsets does not lead to any significant variance reduction. The green triangle in the figure marks the best compromise according to the BIC criterion, without bootstrap, regardless of the number of variables, namely the selection of the triplet $(z_1,z_2,z_3)$. This is exactly to the ground truth. A more quantitative interpretation is offered by the PME-based pie chart in the right side of the figure. Averaged over 100 bootstrap runs, this plot shows the proportion of the output variance accounted for by each variable and the model error. It confirms the large influence of $z_1$, $z_2$, and $x$, followed by a smaller but still non-negligible contribution from $z_3$, and almost no influence from $z_4$ and $z_5$. This distribution closely matches the PME decomposition of the true model.

From a practical point of view, these results indicate that a reliable estimate of $x$ based on $y$ requires at least the measurements of $z_1$ and $z_2$, and ideally also $z_3$. Including $z_4$ and $z_5$, which have no real influence, would only introduce unnecessary noise and degrade performance. This is precisely what is observed in Table \ref{tab:inversion_specic_ase}, which compares estimation results for the three model configurations defined earlier in Section \ref{sec:application}: the \emph{simple} model, which relies only on $y$; the \emph{selected} model, which uses $(z_1, z_2, z_3, y)$; and the \emph{complete} model, which uses all of $(z_1, z_2, z_3, z_4, z_5, y)$.
The table contains several indicators: $R^2$ is the coefficient of determination, ${\text{MAE}}$ is
the mean absolute error, $\mathcal{L}$ is
the length of the 95$\%$ credibility interval, and $\%$ is the proportion of true values lying within that interval. A model is considered effective if it estimates the target variable with low error, narrow credibility intervals, and high $R^2$, while maintaining interval coverage close to 95$\%$. We see here that the \emph{selected} model slightly outperforms the \emph{complete} model , yielding smaller errors and narrower credibility intervals, while the \emph{simple} model fails completely.

\begin{table*}[!ht]
    \centering
    \begin{tabular}{|ccc|GcGc|GcGc|GcGc|}
    \hline
    \multicolumn{3}{|c|}{\textbf{Configurations}} & 
    \multicolumn{4}{c|}{\textbf{Simple model}} &
    \multicolumn{4}{c|}{\textbf{Selected model}} & \multicolumn{4}{c|}{\textbf{Complete model}} \\
    \hline
         $\sigma\m$ & $\rho$ & $\alpha_u$ & $R^2$ & ${\text{MAE}}$ & $\mathcal{L}$ & $\%$ & $R^2$ & ${\text{MAE}}$ & $\mathcal{L}$ & $\%$ & $R^2$ & ${\text{MAE}}$ & $\mathcal{L}$ & $\%$ \\
    \hline
         5 $\%$ & 0.8 & 0 & 0.27 & 0.67 & 4.62 & 96 & 0.93 & 0.19 & 1.08 & 96 & 0.92 & 0.21  & 1.12 & 96   \\
    \hline
    \end{tabular}
    \caption{Inversion results for $(\sigma\m,\rho,\alpha_u)=(5\%,0.8,0)$. The indicators are explained in \ref{subsubsec:Specific_example}}
    \label{tab:inversion_specic_ase}
\end{table*}

\subsubsection{Robustness of the method against noise and correlation}
\label{subsec: Robust noise corr}
This section aims to evaluate the robustness of the proposed method under varying levels of correlation between variables $\rho$ and noise $\sigma\m$. The objective is twofold: to verify whether the influential variables remain accurately identified under these conditions, and to determine the limitations of the method when these factors reach critical thresholds. The case of the unobserved variable $u$ is treated separately in Section \ref{subsec: Robust u}; here we assume $\alpha_u = 0$. To reflect potential discrepancies between the training and deployment environments, we distinguish between two correlation values: $\rho_{\text{train}}$, observed in the training dataset $\mathcal{D}_n$, and $\rho{\text{test}}$, observed during sensor deployment. If $\rho_{\text{train}} = \rho_{\text{test}}$, the calibration and deployment conditions are considered identical. Else, differences between these two values simulate a shift in environmental conditions that can affect inversion performance if inappropriate variables are selected. Note that variable selection is based on the training set only: the effect of modulating $\rho_{\text{test}}$ can only be observed during the inversion phase, as reported in Table \ref{tab:inversion_noise_correlation}.

\begin{table}[!ht]
\centering
\hspace{-1.5cm}
\begin{minipage}{0.495\textwidth}
\centering
\begin{tabular}{|cc|GcGcG|}
\hline
\multicolumn{2}{|c|}{\textbf{Configurations}} & \multicolumn{5}{c|}{\textbf{Variable selection}} \\
\hline
$\sigma\m$ & $\rho=\rho_\text{train}$  & $z_1$ &$z_2$ & $z_3$ & $z_4$ & $z_5$      \\
\hline
\multicolumn{7}{|c|}{\textbf{Low noise and changing correlation}} \\
\hline
2\%& 0   & 100 & 100 & 100 &17&8\\
2\%& 0.5  & 100 & 100 & 100&20 &8\\
2\%& 0.8   &  100& 100 &100 &10 &26\\
2\%& 0.9   & 100 &100  &100 & 7&9\\
\hline
\multicolumn{7}{|c|}{\textbf{Moderate noise and changing correlation}} \\ 
\hline
5\%& 0   & 100 &100  & 100&14 &7\\
5\%& 0.5   & 100 & 100 &100 &15 & 22\\
5\%& 0.8   & 100 & 100 &100 &8 &10\\
5\%& 0.9   &  100& 100 &100 &3 &7\\
\hline
\multicolumn{7}{|c|}{\textbf{High noise and changing correlation}} \\ 
\hline
10\%& 0   & 100 &100  &100 &14 &18\\
10\%& 0.5   &  100& 100 &97 &18 &13\\
10\%& 0.8   &100  & 100 &95 &10 &5\\
10\%& 0.9 & 100 & 100 & 92 &5 &9 \\
\hline
\end{tabular}
\caption*{(a) Influence of the correlation between variables $\rho$ and  and of the noise $\sigma\m$ (in the absence of unmeasured variable)}
\phantomsubcaption
\label{Tab : selection_robust}
\end{minipage}
\hfill
\begin{minipage}{0.495\textwidth}
\centering
\begin{tabular}{|cc|GcGcG|}
\hline
\multicolumn{2}{|c|}{\textbf{Configurations}} & \multicolumn{5}{c|}{\textbf{Variable selection}} \\
\hline
 $\alpha_u$ & $\rho_u$ &$z_1$ &$z_2$ & $z_3$ & $z_4$ & $z_5$ \\
\hline
\multicolumn{7}{|c|}{\textbf{Moderate correlation with $u$}} \\
\hline
 0.1 & 0.8& 100 &100  &100 &17 &25\\
 0.2 & 0.8& 100 &100  &99 &24 &41\\
 0.3 & 0.8& 100 &100  &96 &32 &46\\
\hline
\multicolumn{7}{|c|}{\textbf{Low correlation  with $u$}} \\ 
\hline
0.1 & 0.5& 100 &100  & 100&12 &23\\
0.2 & 0.5& 100 &100  & 97&16 &14\\
0.3 & 0.5&  100& 100 & 79& 14&22\\
\hline
\multicolumn{7}{|c|}{\textbf{No correlation with $u$}} \\
\hline
0.1 & 0& 100 &100  &100 &9 &18\\
 0.2 & 0& 100 &100  &86 &19 &22\\
0.3 & 0& 100 &100  &66 &13 &17\\
\hline
\end{tabular}
\caption*{(b) Varying intensity $\alpha_u$ and correlation $\rho_u$ of the unmeasured variable (for $\rho=5\%$ and $\sigma\m=0.8)$). $\alpha_u=0.1$, $\alpha_u=0.2$, and $\alpha_u=0.3$ correspond to a share of the total variance of $y$ associated with $u$ of 3\%, 10\%, and 23\%, respectively. }
\phantomsubcaption
\label{Tab : selection_robust_u}
\end{minipage}
\caption{Percentage of selection of each variable over 100 bootstrap repetitions, for varying model parameters.}
\label{tab:double_selection}
\end{table}

Table \ref{Tab : selection_robust} first illustrates the effect of noise and correlation on the selection of influential and non-influential variables. At low (2$\%$) and moderate (5$\%$) noise levels, influential variables are correctly selected regardless of correlation. However, at high noise levels (10$\%$), the selection rate of $z_3$ decreases slightly due to its weaker influence compared to the other variables and to the noise. Non-influential variables are rarely selected, with a maximum of 26$\%$. In more complex scenarios (e.g., 10$\%$ noise and a correlation of 0.9), non-influential variables are selected less often, probably due to the simpler models chosen. In summary, the selection method is robust, as it accurately identifies influential variables even in the presence of strong noise and correlation.

\begin{table*}[!ht]
\hspace{-0.75cm}
    \begin{tabular}{|ccc|GcGc|GcGc|GcGc|}
    \hline
    \multicolumn{3}{|c|}{\textbf{Configurations}} & \multicolumn{4}{c|}{\textbf{Simple model}} & \multicolumn{4}{c|}{\textbf{Selected model}} & \multicolumn{4}{c|}{\textbf{Complete model}} \\
    \hline
         $\sigma\m$ & $\rho_\text{train}$ & $\rho_\text{test}$ & $R^2$ & ${\text{MAE}}$ & $\mathcal{L}$ & $\%$ & $R^2$ & ${\text{MAE}}$ & $\mathcal{L}$ & $\%$ & $R^2$ & ${\text{MAE}}$ & $\mathcal{L}$ & $\%$ \\
    \hline
    \multicolumn{15}{|c|}{\textbf{Same environments ($\rho_\text{train}=\rho_\text{test}$) with low noise}} \\
    \hline
         2\%&0&0 & -1.59 & 1.31 & 5.27 & 98 & 0.98 & 0.09 & 0.51 & 98 & 0.98 & 0.09 & 0.52 & 98      \\ 
         2\%&0.5&0.5  & -0.33 & 0.91 & 4.83 & 96  & 0.99 & 0.08 & 0.48 & 98  & 0.98 & 0.09 & 0.52 & 97 \\ 
         2\%&0.8&0.8 & 0.30 & 0.66 & 4.53 & 95  & 0.98 & 0.09 & 0.48 & 96 & 0.98 & 0.09 & 0.49 & 95 \\   
         2\%&0.9&0.9 & 0.53 & 0.53 & 4.36 & 95 & 0.98 & 0.09 & 0.45 & 95 & 0.98 & 0.10 & 0.46 & 94 \\  
    \hline
    \multicolumn{15}{|c|}{\textbf{Same environments ($\rho_\text{train}=\rho_\text{test}$) with intermediate noise}} \\
    \hline
         5\%&0&0&  -1.62 & 1.31 & 5.36 & 98   & 0.91 & 0.21 & 1.16 & 97  & 0.89 &  0.23 &  1.19 & 96  \\  
         5\%&0.5&0.5  & -0.36 & 0.92 & 4.92 & 96 & 0.93 & 0.19 & 1.11 & 98   &  0.92 &  0.20 &  1.13 & 97  \\
         5\%&0.8&0.8 & 0.27 & 0.67 & 4.62 & 96   & 0.93 & 0.19 & 1.08 & 96   &  0.92 &  0.21 &  1.12 & 96   \\    
         5\%&0.9&0.9 & 0.50 & 0.54 & 4.46 & 96  & 0.94 & 0.18 & 0.99 & 97  &   0.94 &  0.19 &  1.00 & 96 \\ 
    \hline
    \multicolumn{15}{|c|}{\textbf{Same environments ($\rho_\text{train}=\rho_\text{test}$) with high noise}} \\
    \hline
         10\%&0&0 & -1.73 & 1.34 & 5.57 & 99 & 0.76 & 0.36 & 2.00 & 97   &0.75 &  0.37 &  1.97 & 97   \\ 
         10\%&0.5&0.5& -0.59 & 1.00 & 4.64 & 97  & 0.80 & 0.34 & 1.85 & 97 &  0.79 &  0.35 &  1.88 & 97  \\
         10\%&0.8&0.8& 0.18 & 0.71 & 4.85& 97  & 0.84 & 0.31 & 1.70 & 97  & 0.84 &  0.31 &  1.65 & 97  \\ 
         10\%&0.9&0.9& 0.42 & 0.59 & 4.69 & 97& 0.87 & 0.28 & 1.54 & 98  &   0.87 &  0.28 &  1.49 & 97 \\
    \hline
    \multicolumn{15}{|c|}{\textbf{Different environments ($\rho_\text{train}\neq\rho_\text{test}$) without correlation ($\rho=0$)}} \\
    \hline
    5\%&0&0.5& -0.59 & 1.00 & 5.50 & 100    & 0.93 & 0.19 & 1.14 & 98  &0.92 & 0.21 & 1.18 & 97  \\ 
    5\%&0&0.8   & -0.10 & 0.82 & 5.58 & 100 & 0.94 & 0.18 & 1.12 & 98   &  0.93 & 0.20 & 1.16 & 98    \\
    5\%&0&0.9   & 0.13 & 0.73 & 5.62 &  100&   0.94 & 0.18 & 1.11 & 98    &0.93 & 0.20 & 1.15 & 97  \\
    \hline  
    \multicolumn{15}{|c|}{\textbf{Different environments ($\rho_\text{train}\neq\rho_\text{test}$) with low correlation ($\rho=0.5$)}} \\
    \hline
    5\%&0.5&0&  -1.20 & 1.19 & 4.82 & 89 &0.90 & 0.22 & 1.13 & 96 & 0.88 & 0.25 & 1.15 & 94\\  
     5\%&0.5&0.8  &0.07 & 0.75&  4.96& 99 &0.94 & 0.18  &1.10& 98& 0.94 & 0.19 & 1.11& 97     \\
     5\%&0.5&0.9   &   0.27 & 0.65&  4.98& 99 & 0.94 & 0.18 & 1.09 &98&  0.94&  0.18&  1.10 &97\\
    \hline
    \multicolumn{15}{|c|}{\textbf{Different environments ($\rho_\text{train}\neq\rho_\text{test}$) with intermediate correlation ($\rho=0.8$)}} \\
    \hline
    5\%&0.8&0&  -0.63 & 1.04 & 4.48 & 83 & 0.84 & 0.27 & 1.05 & 89 & 0.77 & 0.34 & 1.17 & 85\\
    5\%&0.8&0.5&   -0.04 & 0.81&  4.58 &92 &0.91  &0.22 & 1.06& 94 &0.89&  0.24  &1.12 &93\\  
    5\%&0.8&0.9   &  0.42 & 0.59  &4.63 &98& 0.94 & 0.18&  0.99& 97& 0.94&  0.19&  1.00& 96  \\
    \hline
    \multicolumn{15}{|c|}{\textbf{Different environments ($\rho_\text{train}\neq\rho_\text{test}$) with high correlation ($\rho=0.9$)}} \\
    \hline
    5\%&0.9&0&  -0.36 & 0.95 & 4.32 & 80 & 0.82 & 0.30 & 0.95 & 82 &0.75 & 0.36 & 0.99 & 74 \\
    5\%&0.9&0.5&   0.12 & 0.75  &4.42& 88 &0.89&  0.24  &0.97 &90& 0.88 & 0.26 & 0.98& 87\\ 
    5\%&0.9&0.8   &   0.38 & 0.62&  4.45 &93& 0.93&  0.20 & 0.98& 95& 0.92 & 0.21 & 0.99& 93  \\
    \hline
    \end{tabular}
    \caption{Prediction results for varying noise and correlation levels, with identical and different correlation levels in the train and test data sets, in the absence of unmeasured variable. The inversion indicators are explained in \ref{subsubsec:Specific_example}.}
    \label{tab:inversion_noise_correlation}
\end{table*}

We can then use Table \ref{tab:inversion_noise_correlation} to discuss the effect of proper variable selection on the prediction of $x$. We first see that the prediction accuracy for the \emph{simple model} is poor when the correlation is low (0 to 0.5), as reflected by negative $R^2$ values, high MAE, and large credibility intervals. As correlation increases, prediction improves, likely due to compensation between variables, but this improvement does not ensure reliable calibration, especially when the environment changes. In these cases, predictions remain poor, with large credibility intervals and lower percentages of true values within them, highlighting that higher correlation does not lead to robustness across environments.
In contrast, the \emph{selected model}, which focuses on influential variables, provides better accuracy with higher $R^2$, reduced MAE, and smaller credibility intervals, and performs well even when the environment changes. The \emph{complete model}, which includes all variables, does not improve and may even degrade performance due to overfitting, especially in different environmental conditions (when $\rho_\text{train} \neq \rho_\text{test}$).

\subsubsection{Robustness of the method against an unmeasured variable}
\label{subsec: Robust u}

In this section, we assess the effect of an unmeasured variable $u$ on variable selection and prediction performance by modulating  $\alpha_u$ and $\rho_u$. We restrict the analysis to the case where $\sigma\m=5\%$ and $\rho=\rho_{\text{train}}=0.8$. The results in Table \ref{Tab : selection_robust_u} show that the strongly influential variables $z_1$ and $z_2$ are always selected. Although the influence of $z_3$ is weaker ($4\%$ of the total variance), its selection is only slightly sensitive to $u$: the percentage of selection of $z_3$ remains above 79\% in all configurations except for $\rho_u = 0$ and $\alpha_u = 0.3$ (strong influence - 23$\%$ of the variance - of the unmeasured variable, uncorrelated with the influential variables). Even in this degraded configuration, the percentage of selection still reaches $66\%$. Conversely, the non-influential variables $z_4$ and $z_5$ are rarely selected (<25\%) when the correlation level remains moderate ($\rho_u < 0.8$), regardless of $\alpha_u$. The percentage of (wrong) selections of $z_4$ and $z_5$ becomes significant only at high correlation level and moderate to high values of $\alpha_u$. 

The prediction results are presented in Table  \ref{tab:inversion_noise_correlation}. They show that in the same environment ($\rho_\text{train}=\rho_\text{test}$), the \emph{selected model} is comparable to or slightly better than the \emph{complete model}, regardless of $\alpha_u$ and $\rho_u$. Very interestingly, when used in a different environment ($\rho_\text{train}\neq\rho_\text{test}$), while prediction performance deteriorates for both models, the \emph{selected model} significantly outperforms the  \emph{complete model}. Prediction quality remains good overall for the \emph{selected model} despite the unmeasured variable, except in the particularly degraded and quite unrealistic case of a highly influential unmeasured variable that is completely uncorrelated with the influential variables. This further supports the robustness of the proposed algorithm.

\begin{table}[!ht]
    \centering
    \begin{tabular}{|ccccc|GcGc|GcGc|}
    \hline
    \multicolumn{5}{|c|}{\textbf{Configurations}} & \multicolumn{4}{c|}{\textbf{Selected model}} & \multicolumn{4}{c|}{\textbf{Complete model}} \\
     \hline
         $\sigma\m$ & $\rho_\text{train}$& $\rho_\text{test}$ & $\alpha_u$ & $\rho_u$ &$R^2$ & ${\text{MAE}}$ & $\mathcal{L}$ & $\%$&$R^2$ & ${\text{MAE}}$ & $\mathcal{L}$ & $\%$      \\
         \hline
         \multicolumn{13}{|c|}{\textbf{Same environments ($\rho_\text{train}=\rho_\text{test}$)}} \\
         \hline
         5\%& 0.8&0.8 & 0.1 & 0.8&  0.95  &0.17&  1.05& 99& 0.94  &0.19 & 1.06& 97\\
         5\%& 0.8 &0.8& 0.2 & 0.8&  0.93 & 0.21&  1.17& 97&0.92 & 0.22&  1.17 &97 \\
         5\%& 0.8&0.8 & 0.3 & 0.8& 0.91 & 0.24 & 1.32& 97 &0.89 & 0.26 & 1.30& 97\\
         \hline
         5\%& 0.8&0.8 & 0.1 & 0.5& 0.89&  0.24 & 1.13& 97 &0.85 & 0.30&  1.07 &92\\
         5\%& 0.8 &0.8& 0.2 & 0.5& 0.82&  0.34 & 1.53& 95  & 0.77&  0.37&  1.46& 94     \\
         5\%& 0.8&0.8 & 0.3 & 0.5& 0.76&  0.39 & 1.81& 96  & 0.69 & 0.44&  1.75& 92 \\
         \hline       
         5\%& 0.8&0.8 & 0.1 & 0& 0.89  &  0.25&  1.41& 97 &0.88 & 0.26&  1.43& 96\\
         5\%& 0.8&0.8 & 0.2 & 0& 0.80 & 0.35  &1.89 &97 & 0.80&  0.35 & 1.87& 97\\
         5\%& 0.8 &0.8& 0.3 & 0  & 0.74  &0.40 & 2.19& 98 & 0.71 & 0.41 & 2.17& 96 \\
         \hline
         \multicolumn{13}{|c|}{\textbf{Different environments ($\rho_\text{train}\neq\rho_\text{test}$)}} \\
         \hline
         5\% & 0.8 & 0.5 & 0.1 & 0.8 & 0.88& 0.26& 1.07& 91  & 0.85 &0.30& 1.10 &86 \\
         5\% & 0.8 & 0.5 & 0.2 & 0.8 & 0.85& 0.31& 1.09& 83  & 0.76 &0.37& 1.03& 71 \\
         5\% & 0.8 & 0.5 & 0.3 & 0.8 & 0.83&  0.33 & 1.27 &86 & 0.72 & 0.41 & 1.20 &75 \\
         \hline          
         5\% & 0.8 & 0.5 & 0.1 & 0.5 & 0.88& 0.27 &1.11& 87  & 0.80 &0.34 &1.06 &77 \\
         5\% & 0.8 & 0.5 & 0.2 & 0.5 & 0.79& 0.36 &1.50 &89   & 0.71 &0.42 &1.45& 84 \\
         5\% & 0.8 & 0.5 & 0.3 & 0.5 &0.71 & 0.43  &1.77 &90  & 0.56  &0.51 & 1.73& 83 \\
         \hline      
         5\% & 0.8 & 0.5 & 0.1 & 0 &  0.81& 0.33 & 1.32& 87   & 0.80 &0.34& 1.29& 86 \\
         5\% & 0.8 & 0.5 & 0.2 & 0 & 0.70& 0.44& 1.83& 90   & 0.56 &0.50 &1.79& 84    \\
         5\% & 0.8 & 0.5 & 0.3 & 0 &0.56 & 0.53&  2.16& 89  & 0.41  &0.59&  2.07 &85 \\
         \hline

    \end{tabular}
    \caption{Prediction results in the presence of an unmeasured variable, with identical and different correlation levels in the train and test data sets. The inversion indicators are presented in \ref{subsubsec:Specific_example}.}
    \label{tab:selection_u_inversion}
\end{table}

\subsubsection{Conclusion on good selection of variables for sensor robust prediction}

The results in Sections \ref{subsec:application_simulated} through \ref{subsec: Robust u} highlight the critical role of variable selection in ensuring robust and reliable predictions. Interestingly, the presence of an unmeasured variable only moderately interferes with the variable selection process and the subsequent prediction. In all the scenarios, the \emph{selected} model outperforms the \emph{complete model}. It generalizes significantly better in unknown environments, reinforcing its relevance for applications where measurement conditions may deviate from the calibration configuration.

\subsection{Experimental dataset}
\label{subsec:experimentalDataset}

\begin{figure}[!ht]
    \centering
    \includegraphics[scale=1.4]{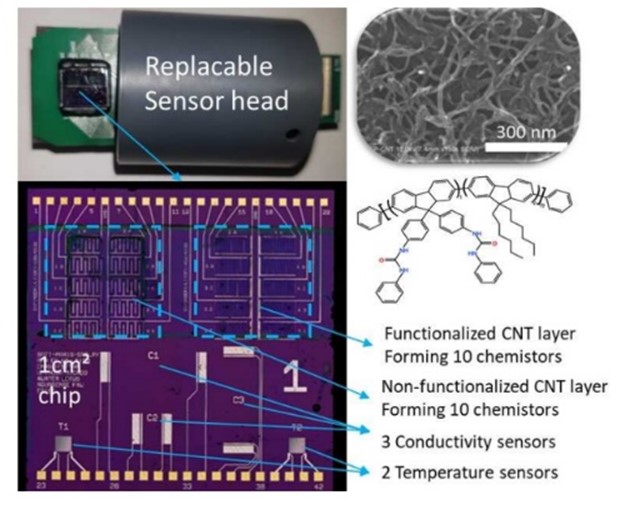}
    \caption{A 10x2 chemiresistor array on a silicon chip (bottom left) fully integrated into a replaceable sensor head (top left). The sensor array consists of ink-jet printed carbon nanotube networks (top right), either unfunctionalized or functionalized with a polyfluorene carrying urea moieties (middle right). 
    }
    \label{fig:Capteur_BL}
\end{figure}

This section evaluates the applicability of the proposed method to calibrate and deploy a carbon nanotube (CNT)-based sensor array under outdoor conditions. We revisit a dataset previously analyzed in \cite{SMAI_Calibration} and \cite{Ozone_CO_IEEE_proce}, where similar calibration techniques were used, but sensor and interferent selection relied on expert judgment and semi-exhaustive parameter search. This data set comes from a 28-day outdoor deployment of a carbon nanotube (CNT)-based chemiresistor array in Marne-La-Vallée, France, near a major traffic artery. As shown in Figure \ref{fig:Capteur_BL}, the sensor array includes 20 chemiresistors — 10 with unfunctionalized CNTs and 10 CNTs functionalized with a polyfluorene polymer. While unfunctionalized CNTs are largely sensitive to various gases and environmental factors (e.g., temperature, humidity), functionalization of CNTs is widely used to increase their selectivity. However, the effectiveness of functionalization strategies under real-world conditions remains unclear due to limited outdoor studies.

The dataset includes time series for six environmental variables (carbon monoxide $CO$, carbon dioxide $CO_2$, nitrogen dioxide $NO_2$, ozone $O_3$, relative humidity (RH), and temperature (Temp)), for six functioning unfunctionalized sensors (labeled 1A, 1B, 1C, 1E, 2A, 2B and 2C), and for five functionalized sensors (labeled 1H, 1G, 1I, 2G, 2H). Some sensors in the array were excluded due to erroneous readings. Due to the high correlation between sensors of the same type, a single representative (1A for unfunctionalized, 1H for functionalized) is used in most analyses. All variables in the dataset are highly correlated (as shown in Figure \ref{fig:Cor_real_data} of Section \ref{Appendix_D}).

The proposed selection method is applied without prior assumptions about sensitivities, treating all variables as potential interferents during calibration. The most influential variables identified during this phase are then targeted for prediction. Our previous study suggested $O_3$ and $CO$ as targets, with RH and temperature as interferents (the previous results are reported in Table \ref{Table.Complexity.Methods} of Section \ref{Appendix_D}). Our goal here is to revisit this previous selection of variables using our new method, hopefully yielding a better understand of the sensors and better predictions. The results are first averaged over 100 random 50/50 splits of dataset, followed by a time-based split to better simulate real-world deployment conditions.

\subsubsection{Derivation of the sensor sensitivities}

\begin{figure}[!ht]
    \centering
    \begin{subfigure}[b]{\textwidth}
        \centering
        \includegraphics[width=0.9\textwidth]{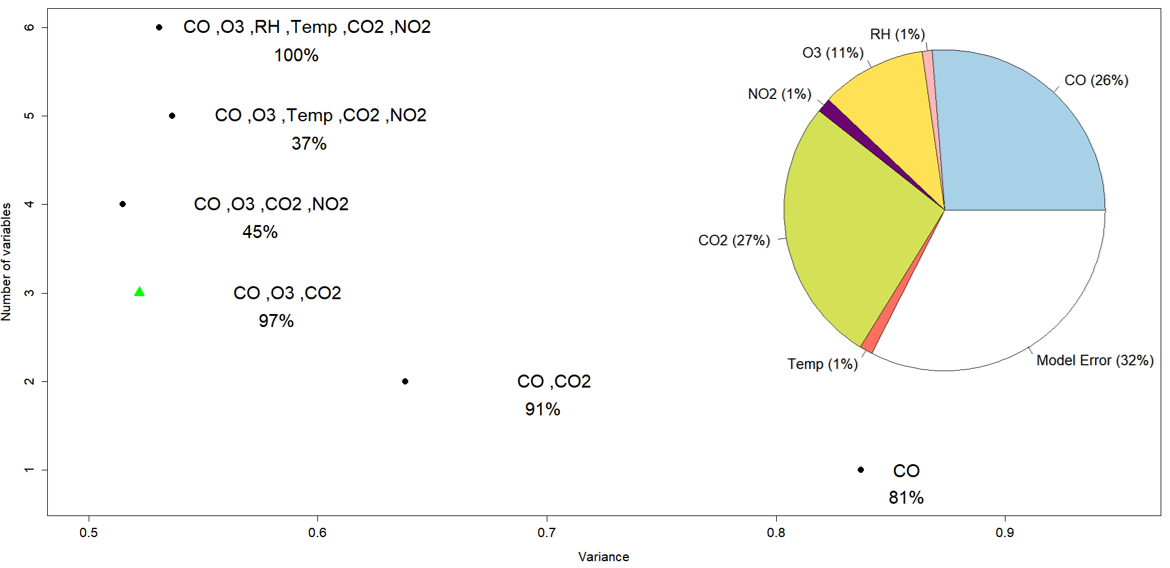}
        \caption{1A - unfunctionalized CNT - sensor}
        \label{Pareto_Front_1A}
    \end{subfigure}
    \begin{subfigure}[b]{\textwidth}
        \centering
        \includegraphics[width=0.9\textwidth]{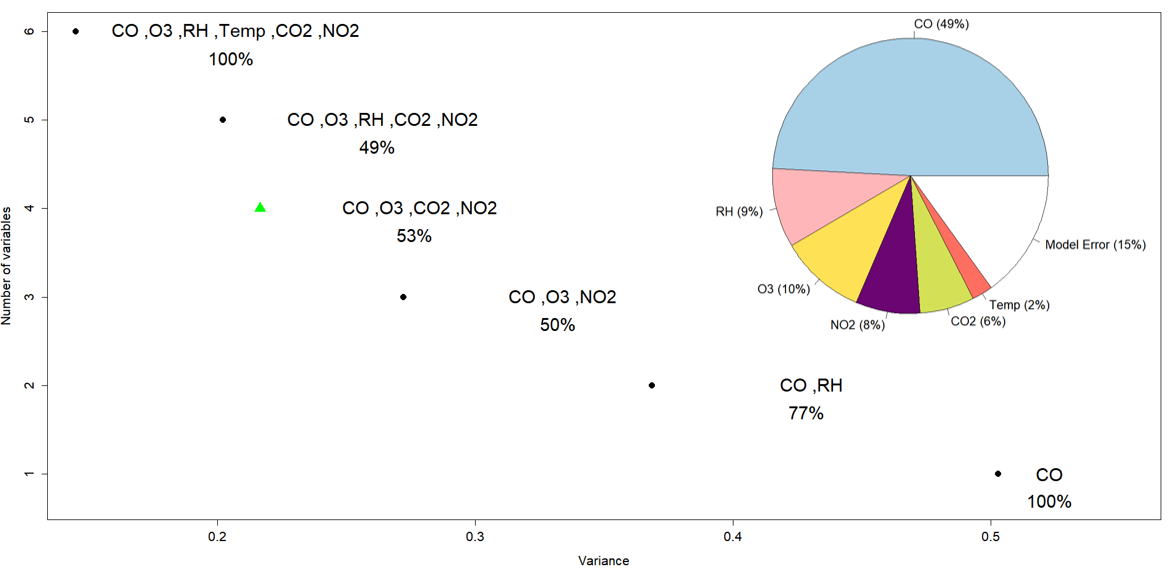}
        \caption{1H - functionalized CNT - sensor}
        \label{Pareto_Front_1H}
    \end{subfigure}
    \caption{Average Pareto front obtained for the 1A - unfunctionalized CNT - sensor (Top) and 1H - functionalized CNT - sensor (Bottom) over 100 random splits of the dataset. The percentage below the selected model represents the percentage of selections of the model. The green triangles represent the \emph{selected models} obtained with the BIC criterion over the 100 bootstrap repetitions. The pie charts represent the mean of the PMEs values over these 100 repetitions.}
    \label{fig:Pareto_front_exp}
\end{figure}

Table \ref{tab : selection1H1A} shows the frequency of variable selection over 100 repetitions, while Figures \ref{Pareto_Front_1A} and \ref{Pareto_Front_1H} display the Pareto fronts and average sensitivity breakdowns for unfunctionalized and functionalized sensors, respectively. (Detailed splits for each optimal model are shown in Figure \ref{fig:PME_selected} of Section \ref{Appendix_D}). We observe that the model consistently includes $CO$, $CO_2$, and $O_3$ for both types of sensors; however, $NO_2$ (strong) and $RH$ (moderate) are only significant for functionalized sensors. This distinction is clearer in the Pareto fronts and pie charts. For the unfunctionalized sensor, one-third of the variance remains unexplained, and the optimal model only includes $CO_2$, $CO$, and $O_3$. These variables explain 32$\%$, 25$\%$, and 9$\%$ of the variance, respectively, with a high model error of 34$\%$. Adding more variables worsens the performance. Overall, this suggests the likely influence of unmeasured variables on the unfunctionalized sensors. These may be volatile organic compounds, which are present in significant concentrations near traffic arteries, are difficult to monitor in real time, and are known to strongly affect CNT sensors. As seen in the simulations, such hidden variables can lead to poor generalization. 

In contrast, the functionalized sensor achieves a much lower model error (15$\%$), suggesting fewer or less influential unmeasured variables and a more reliable model. Here, $CO$ dominates (49$\%$ of the variance), followed by $O_3$, $CO_2$, and $NO_2$ in the optimal four-variable model, which capture 61$\%$, 13$\%$, 6$\%$, and 5$\%$ of the variance, respectively. Although $RH$ appears in two-variable configurations, it is displaced by more informative variables as the model grows, suggesting that it acts as a proxy rather than a true driver. Inclusion of $RH$ and temperature beyond the optimal model adds complexity with limited gain (9$\%$ and 2$\%$ of variance explained), making them unnecessary under current uncertainty levels.

From an application standpoint, comparison of 1H and 1A reveals that functionalization has meaningfully reshaped sensor behavior, greatly enhancing $CO$ sensitivity, while dampening responses to $CO_2$, $NO_2$, and likely other unmeasured variables. Notably, $O_3$ sensitivity remains unaffected. This confirms that functionalization can refine selectivity under real-world conditions.

\begin{table}[!ht]
    \centering
    \begin{tabular}{|c|c|c|c|c|c|c|}
    \hline
          & $CO$ & $O_3$ & $RH$ & $Temp$ & $CO_2$ & $NO_2$\\
          \hline
         1A & 100 & 98 & 22 & 29 & 100 & 38 \\
         \hline
         1H  & 100 & 97 & 73 & 41 & 89 & 95 \\
          \hline
    \end{tabular}
    \caption{Percentage of selection of each variable in the \emph{selected} model for sensors 1H (functionalized CNT) and 1A (unfunctionalized CNT) over 100 random splits of the data set}
    \label{tab : selection1H1A}
\end{table}

\subsubsection{Prediction of the pollutants concentrations}

To further examine sensor sensitivities, each variable in the optimal model was predicted individually, assuming all other variables in the \emph{selected} model known. The results are compared to the predictive performance of the prior (e.g., the prediction using only the environmental variables and not the sensor). The results are summarized in Table \ref{Inv_real_1_variable}, where the columns $R_p^2$ and ${\text{MAE}}_p$ are associated with the prior. (For comparison, the results with the \emph{complete} models are provided in Table \ref{Inv_real_1_variable_complete} of Section \ref{Appendix_D}).

For the unfunctionalized sensor 1A, the predictions show modest $R^2$ values (0.35–0.45) and only slight improvements (10–15$\%$ MAE reduction) over the prior. This limited gain is consistent with the high model error, suggesting modest predictive ability. Conversely, the functionalized sensor 1H yields strong improvements over the prior for $CO$ and $O_3$, and mild ones for $NO_2$ and $CO_2$, suggesting true sensitivity to $CO$ and $O_3$ only. This is further confirmed when studying the prediction of $CO$ and $O_3$ with time-structured splits of the data as shown in Table \ref{tab:pred_tempo_N} of Section \ref{Appendix_D}. 

Assuming that 1H and 1G are sensitive only to $CO$ and $O_3$, these variables were predicted using both sensors and other environmental data in Table \ref{Inv_real_2_variable} of Section \ref{Appendix_D}. The results showed clear improvement in the prediction of $CO$ — comparable to the one-sensor, one-variable case — while the prediction of $O_3$ remained only slightly better than the prior, and similar across sensor combinations. This supports the idea that $O_3$ prediction is mostly derived from correlated variables such as $CO$, rather than direct sensor sensitivity — consistent with PME results indicating that $O_3$ sensitivity is lower than the model error.

For comparison with the literature on $CO$ detection with CNT sensors, the resolution in $CO$ prediction was quantified for the five functionalized CNT sensors (see Figure \ref{fig:resolution_1H} of Section \ref{Appendix_D}). At the $3\sigma$ threshold, the average resolution over the 7 - 7.6 ppm range spans 86 to 98 ppb for the 5 sensors. There are resolution variations over the range: the resolution of the sensor 1H varies between 41 and 144 ppb. These results represent about a tenfold improvement over state-of-the-art laboratory values \cite{LOD_CO_1, LOD_CO_2}, although the concentration ranges studied in the literature are generally much broader (1–1000 ppm). Furthermore, it was found that aggregating data from all five functionalized sensors does not improve prediction accuracy in terms of $R^2$ or MAE, but it does reduce confidence interval lengths by a factor of two, suggesting an effective improvement in resolution when the sensor array is treated as an ensemble  (see Table \ref{tab:5sensors} of Section \ref{Appendix_D}).

In summary, the method allowed exhaustive, unbiased identification of the most informative variables without relying on expert input. It revealed unmeasured interferents, quantified the predictive value for each variable, and demonstrated how CNT functionalization enhanced selectivity to $CO$ while reducing model error, resulting in strong $CO$ predictive capabilities. 

\begin{table}[!ht]
\centering
\begin{tabular}{|c|GcGcG|cccccc|cccccc|cccccc|cccccc|}
\hline
Selected components of $\g{z}$ & $R^2$ & ${\text{MAE}}$ & $\mathcal{L}$ & $R_p^2$ & ${\text{MAE}}_p$  \\ 
\hline
\multicolumn{6}{|c|}{Prediction of $CO$ using sensor 1A} \\
\hline
$O_3,CO_2$ & 0.35&  0.036&0.19& 0.11 &  0.043   \\
\hline
\multicolumn{6}{|c|}{Prediction of $CO$ using sensor 1H} \\
\hline
$O_3,CO_2,NO_2$  & 0.77 & 0.024&  0.12&  0.11 & 0.043 \\
\hline
\multicolumn{6}{|c|}{Prediction of $CO_2$ using sensor 1A} \\
\hline
$O_3,CO$  & 0.45&  5.3& 26 &0.35&  5.8   \\
\hline\multicolumn{6}{|c|}{Prediction of $CO_2$ using sensor 1H} \\
\hline
$O_3,CO,NO_2$ & 0.40&  5.6& 29 &  0.36 & 5.8   \\
\hline
\multicolumn{6}{|c|}{Prediction of $O_3$ using sensor 1A} \\
\hline
$CO,CO_2$ & 0.40 & 5.5 & 30 & 0.26 & 6.2 \\
\hline
\multicolumn{6}{|c|}{Prediction of $O_3$ using sensor 1H} \\
\hline
$CO,CO_2,NO_2$ & 0.66 & 4.0 & 20 & 0.30 & 5.9 \\
\hline
\multicolumn{6}{|c|}{Prediction of $NO_2$ using sensor 1H} \\
\hline
$O_3,CO,CO_2$  & 0.62&  1.3& 5.7 &  0.38 & 1.5   \\
\hline
\end{tabular}
\caption{Separate prediction of $CO$, $CO_2$, $NO_2$ and $O_3$ using either sensors 1H or 1A, and using the \emph{selected} model involving the listed components of $\g{z}$. The columns $R_p^2$ and ${\text{MAE}}_p$ provide the performance of the prior (e.g. of the prediction with only the environmental variables included in the \emph{selected} model). $CO_2$ varies from 409 to 499 {ppm}, $CO$ from 7 to 7.6 ppm, $O_3$ from 15 to 83 ppb, and $NO_2$ varies from 0.7 to 27 ppb.}
\label{Inv_real_1_variable}
\end{table}

\section{Conclusion}
This work presented a variable selection algorithm based on a trade-off between variance minimization and model complexity to improve the predictive capacity of sensors. This paper proposed several tools to post-process the algorithm results for a better understanding of how the sensors work, including the definition of a Pareto front to identify optimal models, and the use of a variance decomposition method for a quantitative analysis of sensor sensitivities. This variable selection method was successfully applied to analytical cases of increasing complexity. The results demonstrated the robustness of the method against several challenges inherent to calibration and sensor deployment in open environments, such as significant measurement noise strong correlations between environmental variables, and the influence of unmeasured variables. The approach was then applied to experimental data obtained from a functionalized carbon nanotube-based sensor array deployed outdoors. The results highlighted how the functionalization of the CNT sensors with a specific polyfluorene polymer bearing urea moieties enabled the sensor to achieve much stronger selectivity to $CO$ than non-functionalized sensors, a much desired goal in the literature. The resolution of the sensors in $CO$ was found remarkable compared to state of the art results obtained in laboratory. 


\bibliographystyle{plain}
\bibliography{biblio}

\appendix

\section{Approximate derivation of the expectation of the conditional variance}
\label{appendixA}

{

Let $\g{X}\m,\g{Z}\m_{\g{\alpha}}$ be the measured values of $\g{x},\g{z}_{\g{\alpha}}$ that we could obtain at any instant that is not included in $\mathcal{D}_n$, and $\g{\alpha}$ be a subset of $\Ac{1,\ldots,d_z}$. In this section, it is important to notice that these two random vectors incorporate two sources of randomness: the first linked to measurement uncertainty, and the second linked to the fact that the values of $\g{x}$ and $\g{z}$ can vary within their definition domain when choosing different times. 
According to Eq. (10), for a given model $\g{f}$ and associated estimators $\widehat{\g{\beta}}$ and $\widehat{\theta}$, the predictor of the measurement of the $m^{\text{th}}$ component of $\g{y}$ that is associated with $\g{X}\m,\g{Z}\m_{\g{\alpha}}$ is defined by:

\Eqa{
\widehat{{v}}(\g{X}\m,\g{Z}\m_{\g{\alpha}}) = \g{f}(\g{X}\m,\g{Z}\m_{\g{\alpha}})^T \g{\widehat{\beta}} + \widehat{\theta}\xi,
}

\noindent{}with $\xi$ a random value that is centered and of variance equal to $1$ (the model error is assumed to be centered, and its variance is likely to be close to $\widehat{\theta}^2$), and which is assumed to be independent of $\g{X}\m,\g{Z}\m_{\g{\alpha}},\widehat{\g{\beta}}$ and $\widehat{\theta}$. We can then decompose:

\Eq{
\begin{split}
\text{Var}& \PP{\widehat{{v}}(\g{X}\m,\g{Z}\m_{\g{\alpha}}) } = \text{Var}\PP{\mathbb{E}\Cr{\widehat{{v}}(\g{X}\m,\g{Z}\m_{\g{\alpha}})\vert \widehat{\g{\beta}},\widehat{\theta}}}
+ \mathbb{E}\Cr{\text{Var}\PP{\widehat{{v}}(\g{X}\m,\g{Z}\m_{\g{\alpha}})\vert \widehat{\g{\beta}},\widehat{\theta}}} \\
& = \mathbb{E}\Cr{\g{f}(\g{X}\m,\g{Z}\m_{\g{\alpha}})}^T\g{C}_{\beta}\mathbb{E}\Cr{\g{f}(\g{X}\m,\g{Z}\m_{\g{\alpha}})}
 + \mathbb{E}\Cr{\widehat{\theta}^2} + \mathbb{E}\Cr{ \g{\widehat{\beta}}^T \text{Cov}(\g{f}(\g{X}\m,\g{Z}\m_{\g{\alpha}})^T \g{\widehat{\beta}}) \g{\widehat{\beta}} }.
\end{split}
}

If we note $\g{m}_{f}=\mathbb{E}\Cr{\g{f}(\g{X}\m,\g{Z}\m_{\g{\alpha}})}$ and $\g{C}_f=\text{Cov}(\g{f}(\g{X}\m,\g{Z}\m_{\g{\alpha}}))$, and if $\text{Tr}\PP{\cdot}$ stands for the trace operator, it comes:

\Eq{
\text{Var} \PP{\widehat{{v}}(\g{X}\m,\g{Z}\m_{\g{\alpha}})} = \g{m}_{f}^T\g{C}_{\beta}\g{m}_{f} + \mathbb{E}\Cr{\widehat{\theta}^2}+ \g{m}_{\beta}^T\g{C}_f\g{m}_{\beta} + \text{Tr}\PP{\g{C}_f\g{C}_\beta}.
}

Noticing that

\Eq{
\g{m}_{\beta}^T\g{C}_f\g{m}_{\beta} = \text{Var}\PP{\mathbb{E}\Cr{ \widehat{{v}}(\g{X}\m,\g{Z}\m_{\g{\alpha}})\vert \g{X}\m,\g{Z}\m_{\g{\alpha}} }},
}

\noindent{}it finally comes:

\Eq{
\begin{split}
\mathbb{E} & \Cr{ \text{Var}\PP{ \widehat{{v}}(\g{X}\m,\g{Z}\m_{\g{\alpha}})\vert \g{X}\m,\g{Z}\m_{\g{\alpha}} }} \\
& = \text{Var} \PP{\widehat{{v}}(\g{X}\m,\g{Z}\m_{\g{\alpha}})} - \text{Var}\PP{\mathbb{E}\Cr{ \widehat{{v}}(\g{X}\m,\g{Z}\m_{\g{\alpha}})\vert \g{X}\m,\g{Z}\m_{\g{\alpha}} }} \\
& = \g{m}_{f}^T\g{C}_{\beta}\g{m}_{f} + \mathbb{E}\Cr{\widehat{\theta}^2} + \text{Tr}\PP{\g{C}_f\g{C}_\beta}.
\end{split}
}

}

{

\section{Derivation of the expressions of $\g{m}$ and $\g{C}$}
\label{appendix_B}

Let $f_{\g{x}_\star\vert \g{y}\m_\star,\g{z}\m_\star}$ be the posterior PDF of $\g{x}_\star$, that is to say 
the PDF of $\g{x}_\star$ given the observations of $\g{y}\m_\star$ and $\g{z}\m_\star$. Using the Bayes theorem, this PDF can be rewritten as

\Eq{ f_{\g{x}_\star\vert \g{y}\m_\star,\g{z}\m_\star}(\g{x}) = \widetilde{c}\times L(\g{x}) f_{\g{x}_\star}(\g{x}), \ \ \g{x}\in\R^{d_x},}

\noindent{}where $\widetilde{c}$ is a normalization constant, and $L$ is the likelihood function associated to the PDF of $\g{y}\m_\star\vert \g{x}_\star, \g{z}\m_\star$. Assuming now that the measurement error $\g{\varepsilon}_\star^x=\g{x}\m_\star-\g{x}_\star$ is small, the expression given in Eq. (17) can be linearized to obtain the following relationship between $\g{y}\m_\star$ and $\g{x}_\star$:

\Eq{
\begin{split}
    \g{y}\m_\star & \approx  \widehat{\g{\Theta}}\g{\xi} + \g{F}(\g{x}_\star, \g{z}\m_\star)\widehat{\g{\beta}}^{\text{tot}} \\ & + \sum_{j=1}^{d_x}\frac{\partial \g{F}}{\partial x_j}(\g{x}_\star, \g{z}\m_\star)\widehat{\g{\beta}}^{\text{tot}}\varepsilon^x_{j,\star}.
\end{split}
}

Reminding that the vector $\g{\xi}$ is modelled by a centered random vector with a covariance matrix equal to the identity matrix, and assuming, first, that 

\Eq{\mathbb{E}\Cr{\g{\varepsilon}_{\star}^x }=\g{0}, \ \ \text{Cov}\PP{\g{\varepsilon}_{\star}^x }=\text{diag}\PP{\sigma_1^2,\ldots,\sigma_{d_x}^2},}

\noindent{}and, secondly, that the random quantities $\g{\xi},\widehat{\g{\beta}}^{\text{tot}},\varepsilon^x_{j,\star}$ are statistically independent, we can then calculate:

\Eq{ \mathbb{E}\Cr{ \g{y}\m_\star\vert \g{x}_\star, \g{z}\m_\star  } = \g{m}(\g{x}_\star)=\g{F}(\g{x}_\star, \g{z}\m_\star)\mathbb{E}\Cr{\widehat{\g{\beta}}^{\text{tot}}},\label{formMoy}}

\Eq{
\begin{split}
\text{Cov} & (\g{y}\m_\star \vert \g{x}_\star, \g{z}\m_\star )= \g{C}(\g{x}_\star)=\mathbb{E}\Cr{\widehat{\g{\Theta}}\widehat{\g{\Theta}}^T} \\ & + \g{F}(\g{x}_\star, \g{z}\m_\star) \text{Cov}\PP{\widehat{\g{\beta}}^{\text{tot}}} \g{F}(\g{x}_\star, \g{z}\m_\star)^T \\
& + \sum_{j=1}^{d_x}\sigma_j^{2}\frac{\partial \g{F}}{\partial x_j}(\g{x}_\star, \g{z}\m_\star) \left\{\text{Cov}\PP{\widehat{\g{\beta}}^{\text{tot}}} \right. \\
& \hspace{0.5cm} \left. + \mathbb{E}\Cr{\widehat{\g{\beta}}^{\text{tot}}}\mathbb{E}\Cr{\widehat{\g{\beta}}^{\text{tot}}}^T\right\}
 \PP{\frac{\partial \g{F}}{\partial x_j}(\g{x}_\star, \g{z}\m_\star)}^T.
\end{split}\label{formeCov}
}

As proposed in \cite{SMAI_Calibration}, if we now approximate the PDF of $\g{y}\m_\star\vert \g{x}_\star, \g{z}\m_\star$ by a Gaussian PDF, we obtain for the posterior PDF of $\g{x}_\star$ the function that is defined in Eq. (18). 

\medskip

Looking at Eqs. (\ref{formMoy}) and (\ref{formeCov}), we see that functions $\g{m}$ and $\g{C}$ only depend on $\mathbb{E}\Cr{\widehat{\g{\beta}}^{\text{tot}}}$, $\text{Cov}\PP{\widehat{\g{\beta}}^{\text{tot}}}$ and $\mathbb{E}\Cr{\widehat{\g{\Theta}}\widehat{\g{\Theta}}^T}$, which can again be approximated using a bootstrap procedure as it is done in \ref{appendixA}.

}

\section{Definition of the PME indices}
\label{appendix_C}

Let 

$$h: \left\{ \begin{array}{ccc}
\R^d & \rightarrow & \R \\
\g{w} & \mapsto & h(\g{w})
\end{array} \right.
$$ 

\noindent{}be a real-valued function defined on $\R^d$ with $d>1$, and $\g{W}$ be a $d-$dimensional random vector. The computation of the PME indices of $h(\g{W})$ follows the work achieved in \cite{PME}. For each $1\leq j\leq d$, the PME value of $h(\g{W})$ associated with the $j^{\text{th}}$ component of $\g{W}$ is noted $\delta_j$, and using the notations of \cite{PME}, it is defined by:

\Eq{\delta_j=\sum_{\pi\in\mathcal{S}_D}p(\pi)\Cr{S^T\PP{C_{\pi(j)}(\pi)}-S^T\PP{C_{\pi(j)-1}(\pi)}},}

\noindent{}where:

\BI{
\item $D=\Ac{1,\ldots,d}$ is the set of all possible indices of $\g{W}$,
\item $\mathcal{S}_D$ is the set of all permutations of $D$,
\item $\pi=(\pi_1,\ldots,\pi_d)$ is a particular element of $\mathcal{S}_D$,
\item $\pi(j)$ is the position of $j$ in $\pi$ such that $\pi_{\pi(j)}=j$,
\item $C_j(\pi)=\Ac{\pi_k: \ k\leq j}$ is the set of the $j^{\text{th}}$ first indices in the ordering $\pi$ with the convention that $C_0(\pi)=\emptyset $,
\item for any subset of indices $A\subset D$, $\g{W}_A$ is the random vector gathering the components of $\g{W}$ whose indices are in $A$, and $S^T(A)$ is the total Sobol index associated with $A$ (see \cite{Sobol} for more details about Sobol indices) so that:

\Eq{ S^T(A) = 1 - \frac{\text{Var}(\mathbb{E}\Cr{h(\g{W})\ \vert \ \g{W}_A})}{\text{Var}(h(\g{W}))}, }

\item $p(\pi)$ is a probability mass function defined as

\Eq{p(\pi)=\frac{L(\pi)}{\sum_{\sigma \in\mathcal{S}_D}L(\sigma)}, \ \ L(\pi)=\PP{\prod_{k\in D}S^T(C_k(\pi))}^{-1}.}

}

These indices, which require the calculation of a large number of conditional expectation variances, are calculated in practice by pick-freeze sampling methods \cite{da2021basics}, in exactly the same way as Sobol indices are classically calculated, without this being a real numerical challenge, given that the execution time of the prediction model is extremely fast.

\section{Resolution analysis}
\label{Appendix_D}

Using the previous notations, an additional post-processing of the optimal model is proposed: it aims to compute the \emph{resolution} of the sensor to its targets $\g{x}$ and interferents $\g{z}_{\g{\alpha}}$. For simplicity, we no longer distinguish between $\g{x}$ and $\g{z}_{\g{\alpha}}$ in this part, and we group the possible values of $\g{x}$ and $\g{z}_{\g{\alpha}}$ at a time that is not in $\mathcal{D}_n$ in the $q$-dimensional vector $\g{W}=(\g{X},\g{Z}_{\g{\alpha}})$, with $q=d_x+\sharp \g{\alpha}$. The model predicting the measured value of the $m^{\text{th}}$ component of $\g{y}$ is then written

\Eq{\widehat{v}(\g{W})=\g{f}(\g{W})^T\widehat{\g{\beta}}+\widehat{\theta}\xi.}

With these notations, the \emph{average} resolution of level $k$ of the sensor $m$ at the value $w_j$ for the $j^{\text{th}}$ component of $\g{W}$ can then be calculated as the minimum fluctuation $\delta_j(w_j)$ that is likely to cause a sensor response variance that is greater than $k$ times the model error, i.e. the fluctuation such that:

\Eq{\delta_j(w_j) = \min \Ac{\delta\geq 0: \ \mathbb{E}_{\g{W}_{-j}}\Cr{\text{Var}_\zeta\PP{\widehat{v}\PP{\g{W}_{-j},w_j+\delta \times \zeta}  \vert \g{W}_{-j}} \geq k\times \widehat{\theta} }},}

\noindent{}where $\g{W}_{-j}$ is the vector containing all the components of $\g{W}$ except the $j^{\text{th}}$ one, and where $\zeta\sim\mathcal{N}(0,1)$ is a standard Gaussian random variable. This sensor resolution is said to be \emph{averaged}, since we are interested in the average over all possible values of $\g{W}_j$ of the sensor variance that is due to $W_j$ when  we focus on values around $w_j$. This resolution is defined as a function of $w_j$, such that its smallest values indicate the ranges where the sensor is most sensitive to $W_j$, and its largest values indicate the ranges where the sensor is least sensitive to $W_j$. 
To calculate this resolution, we propose to use a Gaussian variable $\zeta$, but any other random variable with mean zero and variance 1 could equally well be used (without greatly affecting the results).

\begin{figure}[!ht]
    \centering
    \includegraphics[scale=0.5]{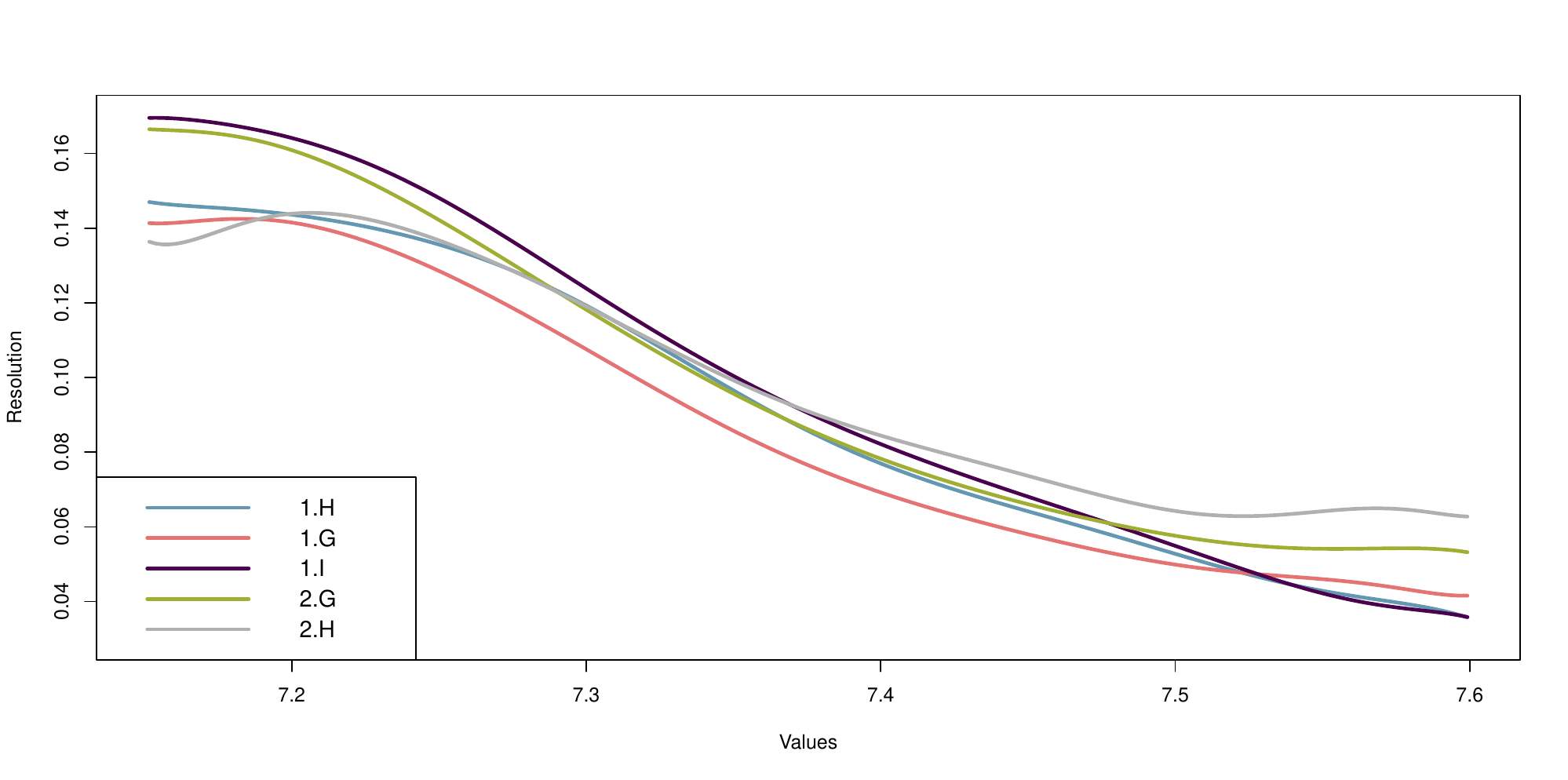}
    \caption{$3\sigma$-resolution (in ppm) in $CO$ for the 1H-type sensors (e.g., $k=3$).}
    \label{fig:resolution_1H}
\end{figure}







\section{Additional figures and tables regarding the application on the experimental dataset}

\begin{figure}[!ht]
    \centering
    \includegraphics[scale=0.5]{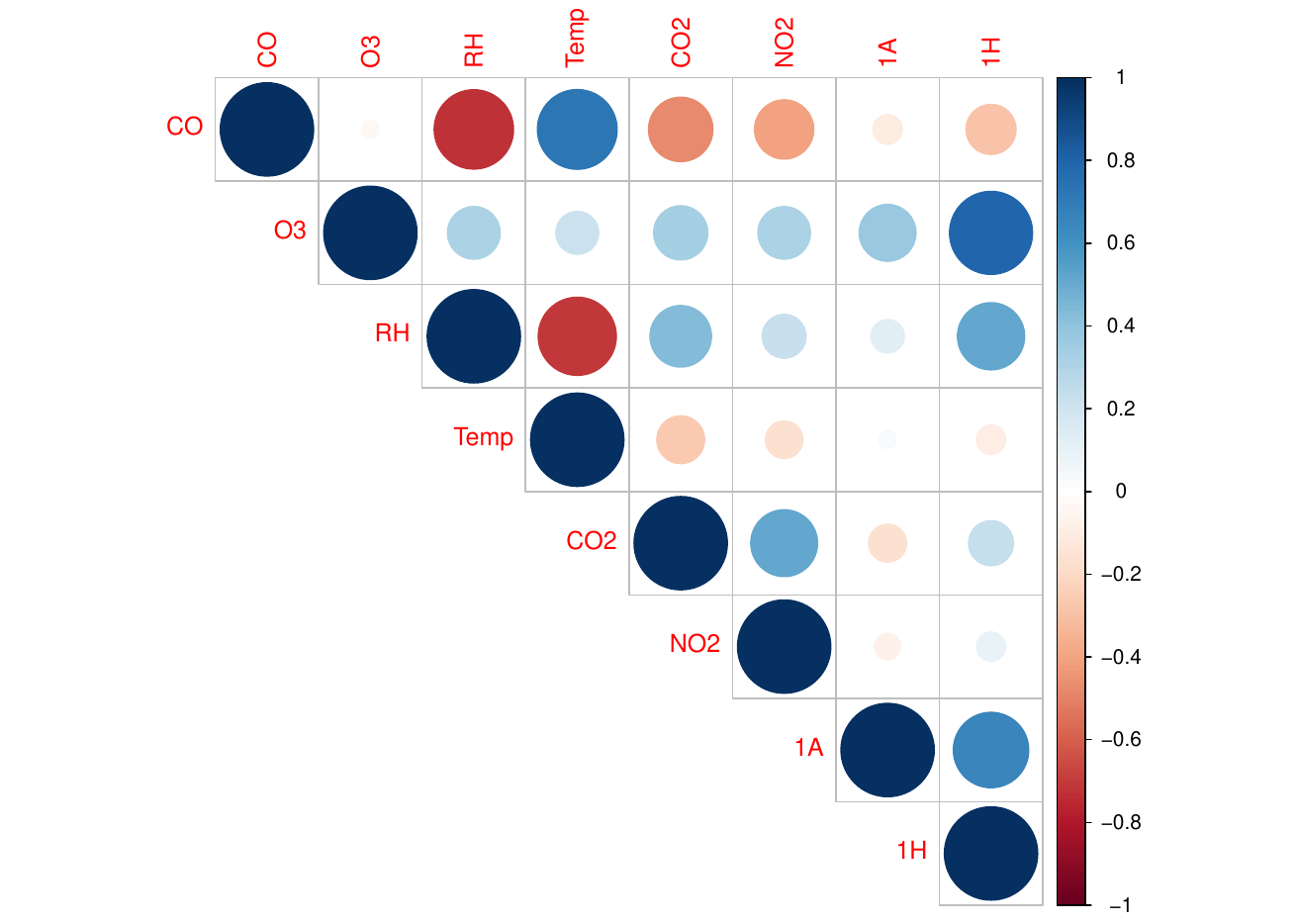}
    \caption{Statistical correlations on the real dataset for the environmental variables and the two sensor outputs. }
    \label{fig:Cor_real_data}
\end{figure}

\begin{table}[!ht]
  \begin{tabular}{cccccc}
$R^2_1$ & $R^2_2$& $\text{MAE}_1$  & $\text{MAE}_2$ & $\mathcal{L}^{90\%}_1$  & $\mathcal{L}^{90\%}_2$ \\
\hline
0.55 $\pm$ 0.072 & 0.74 $\pm$ 0.020 & 5.1 $\pm$ 0.19 &0.030 $\pm$ 0.0022 &18 $\pm$ 1.34 & 0.083 $\pm$ 0.011
\end{tabular}
 \caption{Prediction with GLR 
 of $CO$ and $O_3$ with sensors 1H and 1A, and with temperature and relative humidity as interferents. The index 1 is for $O_3$ (in ppb) while the index 2 is for $CO$ (in ppm). The values correspond to the empirical averages and standard deviations of the indicators obtained over 5 random splits of the dataset. The length of the credibility intervals was computed at 90\%.}
  \label{Table.Complexity.Methods}
\end{table}

\begin{figure}[!ht]
\hspace{-2cm}
\begin{subfigure}{0.45\linewidth}
    \includegraphics[scale=0.55]{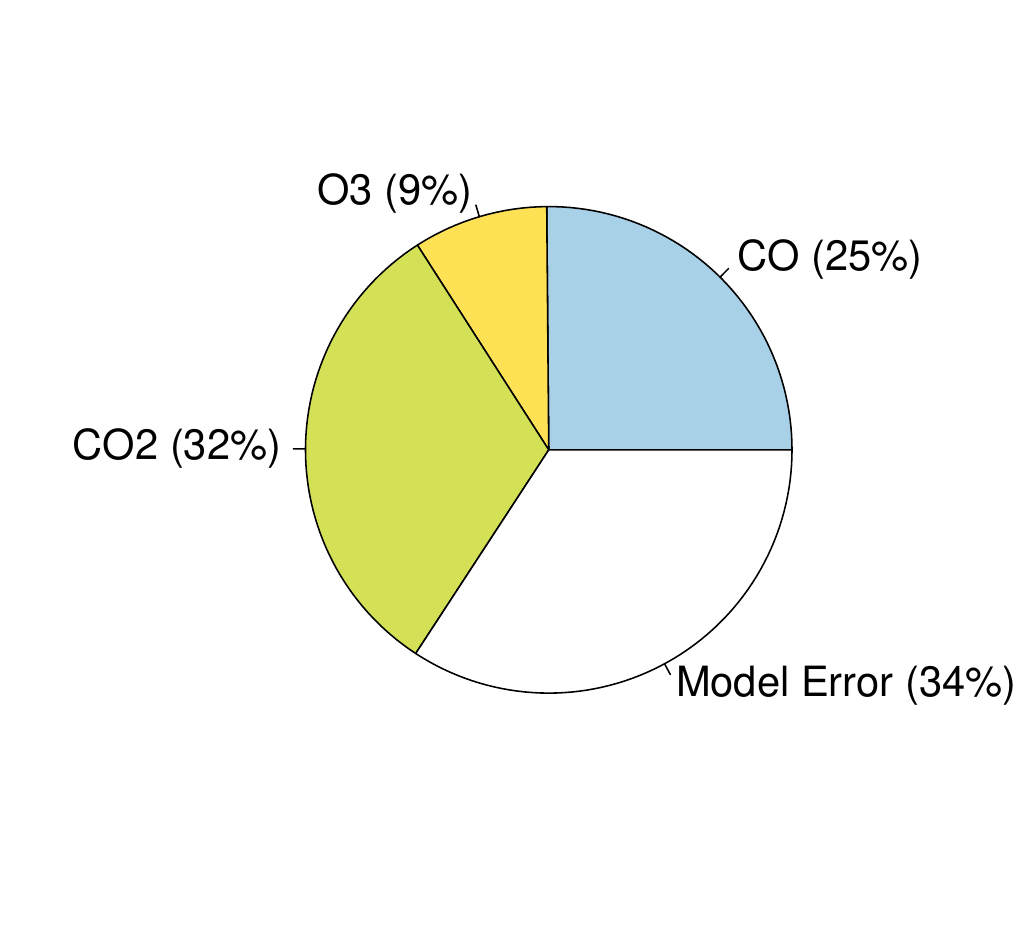}
    \caption{PME obtained for the selected model for the 1A sensor}
    \label{fig:PME_1A_selected}
\end{subfigure}
\hspace{1cm}
\begin{subfigure}{0.45\linewidth}
    \includegraphics[scale=0.55]{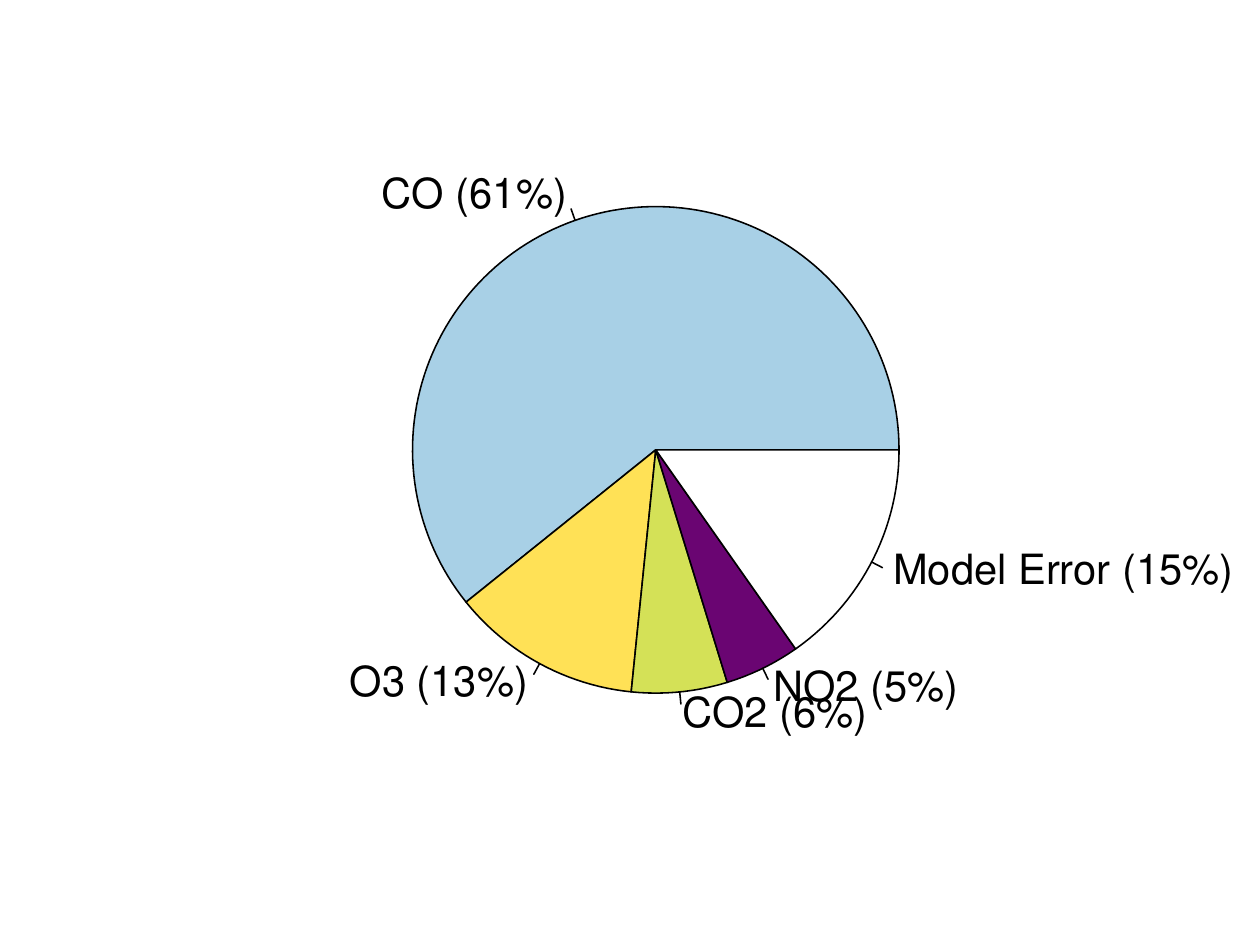}
    \caption{PME obtained for the selected model for the 1H sensor}
    \label{fig:PME_1H_selected}
\end{subfigure}
\caption{PME computed on the selected model for 1A and 1H sensors}
\label{fig:PME_selected}
\end{figure}


\begin{table}[!ht]
\centering
\begin{tabular}{|c|GcGcG|cccccc|cccccc|cccccc|cccccc|}
\hline
\multicolumn{6}{|c|}{Prediction of $CO$ using sensor 1A} \\
\hline
Selected components of $\g{z}$ & $R^2$ & $MAE$ & $\mathcal{L}$ & $R_p^2$ & $MAE_p$  \\ 

\hline
$O_3,CO,NO_2,RH,Temp$ & 0.74 &0.021 &0.11  &0.67 &0.023  \\
\hline
\multicolumn{6}{|c|}{Prediction of $CO$ using sensor 1H} \\
\hline
$O_3,CO_2,NO_2,RH,Temp$ & 0.88 & 0.016&  0.072&   0.67&  0.023 \\
\hline
\multicolumn{6}{|c|}{Prediction of $CO_2$ using sensor 1A} \\
\hline
$O_3,CO,NO_2,RH,Temp$  & 0.66 & 4.0& 20  & 0.62&  4.1 \\
\hline\multicolumn{6}{|c|}{Prediction of $CO_2$ using sensor 1H} \\
\hline
$O_3,CO,NO_2,RH,Temp$ & 0.68 &3.8 & 21 & 0.62&4.1  \\
\hline
\multicolumn{6}{|c|}{Prediction of $O_3$ using sensor 1A} \\
\hline
$CO,CO_2,NO_2,RH,Temp$ & 0.83    &2.9 & 14 & 0.82 & 3.0 \\
\hline
\multicolumn{6}{|c|}{Prediction of $O_3$ using sensor 1H} \\
\hline
$CO,CO_2,NO_2,RH,Temp$ & 0.84 &2.8 &14& 0.82& 3.0 \\
\hline
\multicolumn{6}{|c|}{Prediction of $NO_2$ using sensor 1H} \\
\hline
$O_3,CO,CO_2,RH,Temp$  & 0.69 &1.1 & 4.4 &0.65 &1.2  \\
\hline
\end{tabular}
\caption{Separate prediction of $CO$, $CO_2$, $NO_2$, and $O_3$ using sensors 1H or 1A and the \emph{Complete} model (e.g. with all the available environmental variables). $CO_2$ varies from 409 to 499 ppm, $CO$ from 7 to 7.6 ppm, $O_3$ from 15 to 83 ppb, and the $NO_2$ varies from 0.7 to 27 ppb.}
\label{Inv_real_1_variable_complete}
\end{table}

\begin{table}[!ht]
\centering
\begin{tabular}{|c|GcGcG|cccccc|cccccc|cccccc|cccccc|}
\hline
\multicolumn{6}{|c|}{Prediction of $CO$ using 1H (selected model)} \\
\hline
Scenarios & $R^2$ & $MAE$ & $\mathcal{L}$ & $R_p^2$ & $MAE_p$  \\ 
\hline
1 &0.60 & 0.032 & 0.10 & -0.63 & 0.060 \\
2 & 0.54 & 0.036 & 0.10 & -0.62 & 0.064 \\
3 &0.028 & 0.034 & 0.12 & -2.81 & 0.066 \\
\hline
\multicolumn{6}{|c|}{Prediction of $O_3$ using 1H (selected model)} \\
\hline
1 & 0.47 & 5.8 & 20 & 0.19 & 7.1 \\
2 & 0.26 & 6.9 & 20 & -0.002 & 8.0 \\
3 & 0.33 & 7.5 & 26 & -0.017 & 9.3 \\
\hline
\end{tabular}
\caption{Separate prediction of $CO,CO_2$ and $O_3$ using 1H and 1A sensors with the \emph{selected} models. The predictions are made according to three scenarios: 1) calibration over all the even days and prediction over all the odd days, or 2) calibration over alternating pairs of days (e.g., calibration over days 1,2,5,6,9,10... and prediction over days 3,4,7,8), or 3) calibration over the first and last 10 of the dataset and prediction over the middle 7 days.}
\label{tab:pred_tempo_N}
\end{table}

\begin{table}[!ht]
\centering
\begin{tabular}{|c|GcGcG|cGcGc|cccccc|cccccc|cccccc|}
\hline
 & \multicolumn{5}{|c|}{Prediction of $CO$} & \multicolumn{5}{|c|}{Prediction of $O_3$} \\
\hline
Selected components of $\g{z}$ & $R^2$ & ${\text{MAE}}$ & $\mathcal{L}$ & $R_p^2$ & ${\text{MAE}}_p$ &$R^2$ & ${\text{MAE}}$ &
$\mathcal{L}$ & $R_p^2$ & ${\text{MAE}}_p$  \\ 
\hline
\multicolumn{11}{|c|}{Prediction of $CO$ and $O_3$ using sensors 1H and 1A } \\
\hline
$CO_2, NO_2$   & 0.65 & 0.029 & 0.19 & -0.03 & 0.048 & 0.29 & 6.1 & 30 & 0.14 & 7.1\\
\hline
\multicolumn{11}{|c|}{Prediction of $CO$ and $O_3$ using sensors 1H and 1G } \\
\hline
$CO_2,NO_2$  & 0.64& 0.030 & 0.19  &-0.03   & 0.048 &0.27   &6.0 & 24  & 0.14    & 7.1  \\
\hline
\end{tabular}
\caption{Joint prediction of $CO$ and $CO_2$ using different sensors combinations. $CO$ varies from 7 to 7.6 ppm, $O_3$ from 15 to 83 ppb.} 
\label{Inv_real_2_variable}
\end{table}

\begin{table}[!ht]
\centering
\begin{tabular}{|c|GcGcG|cccccc|cccccc|cccccc|cccccc|}
\hline
Selected components of $\g{z}$ & $R^2$ & $MAE$ & $\mathcal{L}$ & $R_p^2$ & $MAE_p$  \\
\hline
\multicolumn{6}{|c|}{Prediction of $CO$ using 5 sensors of the same type as 1H} \\
\hline
$O_3,CO_2,NO_2$ & 0.71 & 0.026&  0.072&  0.11 & 0.043 \\
\hline
\multicolumn{6}{|c|}{Prediction of $O_3$ using sensor 5 sensors of the same type as 1H} \\
\hline
$CO,CO_2,NO_2$ & 0.60 & 5.0&  15   &  0.30 & 5.9 \\
\hline

\end{tabular}
\caption{Prediction of $CO$ and $O_3$ separately using all the sensors of type $1H$}
\label{tab:5sensors}
\end{table}

\end{document}